\documentclass[pre,amsmath,showpacs]{revtex4}
\usepackage{times}
\usepackage{graphics}
\usepackage{graphicx}
\usepackage{epsf}
\usepackage{epsfig}

\newcommand{\be}{\begin{equation}}
\newcommand{\ee}{\end{equation}}
\newcommand{\BE}{\begin{eqnarray}}
\newcommand{\EE}{\end{eqnarray}}
\newcommand{\BEn}{\begin{eqnarray*}}
\newcommand{\EEn}{\end{eqnarray*}}
\newcommand{\barr}{\begin{array}} 
\newcommand{\earr}{\end{array}}
\newcommand{\bit}{\begin{itemize}}      
\newcommand{\eit}{\end{itemize}}
\newcommand{\bfl}{\begin{flusleft}}
\newcommand{\efl}{\end{flusleft}}
\newcommand{\bfr}{\begin{flushright}}
\newcommand{\bc}{\begin{center}}
\newcommand{\ec}{\end{center}}
\newcommand{\ben}{\begin{enumerate}}    
\newcommand{\een}{\end{enumerate}}
\newcommand{\cl}{\centerline}

\begin{document}

\title{Thermally activated processes in polymer dynamics}

\author{Lorenzo Bongini$^{(1)}$, Roberto Livi$^{(2),(1)}$, 
Antonio Politi$^{(3),(1)}$, and Alessandro Torcini$^{(3),(1)}$} 

\affiliation{$(1)$ INFM, UdR Firenze, via Sansone, 1 - 
I-50019 Sesto Fiorentino, Italy\\
$(2)$ Dipartimento di Fisica, Universit\'a di Firenze, 
via Sansone, 1 - I-50019 Sesto Fiorentino, Italy\\
$(3)$ Istituto Nazionale di Ottica Applicata,
L.go E. Fermi, 6 - I-50125 Firenze, Italy}

\date{\today}

\begin{abstract}
Jumps between neighboring minima in the energy landscape of both homopolymeric
and heteropolymeric chains are numerically investigated by determining the
average escape time from different valleys. The numerical results are compared
to the theoretical expression derived by Langer [J.S. Langer, Ann. Phys.
{\bf 54} (1969) 258] with reference to a $2N$-dimensional space. Our simulations
indicate that the dynamics within the native valley is well described by a
sequence of thermally activated process up to temperatures well above the
folding temperature. At larger temperatures, systematic deviations from the
Langer's estimate are instead observed. Several sources for such discrepancies
are thoroughly discussed.
\end{abstract}
\maketitle


\section{Introduction}

Polymeric chains exhibit quite a rich variety of dynamical properties. At high
temperatures, kinetic energy is large enough to allow a chain exploring most
of the accessible phase-space. In this regime, the polymer typically assumes a
``random coil'' structure. At intermediate temperatures, internal forces and
the interaction with the solvent become strong enough to stabilize compact
configurations~\cite{degennes}. 
However, kinetic energy fluctuations are still able to drive the chain from one
to another minimum of the energy landscape. The properties of this itinerant
dynamics depend on several factors: the height of the barriers separating
neighboring minima, their accessiblity and, more generally, the overall
structure of the energy landscape. Upon further decreasing the temperature, an
heteropolymer typically undergoes a glass transition and may freeze in one of
several distinct free-energy minima. Only some peculiar heteropolymers exhibit
a transition to a ``folding regime'', i.e. are characterized by a relatively
fast convergence towards the absolute energy minimum, irrespectively of the
initial state. In this case, the heteropolymer is said to be a ``good folder''
and it can be viewed as a specimen of a protein, that always evolves to its
native configuration (NC) ~\cite{crei}.

Independently whether a given polymer is homogeneous or heterogeneous, whether
it is a good or a bad folder, a complete understanding of its dynamical
properties passes through the description of the jump processes between
different energy valleys \cite{still2,berry0}. Free-energy valleys are indeed
collections of distinct minima and studying the connectivity of such minima
can help identifying and parametrizing the relevant macroscopic states
~\cite{krivov,evans}. In the hope to eventually make substantial progress 
along this line, in this paper we
aim at testing the validity of the expressions utilized to characterize the
single escape processes. In the current literature, the escape is often viewed
as an activation process and Kramers-like formulae, derived for low--dimensional
systems, are applied to characterize high--dimensional systems without testing
their validity. In this paper we present a detailed check of the formula derived
by Langer in 1969 \cite{langer}, finding that the escape process is strongly
influenced by the entropic contribution associated with the local geometry of
the energy landscape.

In Sec. 2 we introduce the polymer model used as a testing ground for the
numerical analysis of activation processes in relatively high-dimensional
systems~\cite{still}. It consists of a chain of two types of beads embedded in a
two-dimensional space. In the same section, we briefly recall the relevant
properties of both homogeneous and heterogeneous systems upon varying the
temperature. In Sec. 3, the general theoretical ideas lying behind the
derivation of Langer's formula are briefly summarized. The technical
difficulties associated with the determination of geometrical factors are also
discussed together with some possible approximation schemes. In Section 4,
theoretical predictions are compared with numerical simulations for specimens
of bad and good folders. In spite of an overall qualitative agreement, 
systematic deviations are found at relatively high temperatures, the origin of
which is discussed in Sec. 5, where several effects are separately discussed.
Finally, in Sec. 6, the main conclusions are summarized and the open problems
briefly recalled.

\section{The model: definitions and thermodynamical properties}

In this paper we study the escape process from an energy valley with reference
to a model thoroughly investigated in \cite{tlp}, where the authors slightly
modified a previous version, originally introduced in~\cite{still}. The model,
designed to simulate sequences of amino acids interacting within a solvent,
describes a chain of $L$ monomers embedded in a 2-dimensional space.  At
variance with \cite{still}, where monomers were rigidly linked along the
backbone, in \cite{tlp} a nearest-neighbor harmonic potential was instead
assumed,
\begin{equation}
V_1 (r_{i,i+1}) = \alpha (r_{i,{i+1}}-r_0)^2  ,
\label{v1}
\end{equation}
where $r_{i,j} = \sqrt{(x_i-x_j)^2 +(y_i-y_j)^2}$ is the distance between the
$i$th and the $j$-th monomer, while $x_i$, $y_i$ are the coordinates of the
$i$th monomer. Without loss of generality, the equilibrium distance $r_0$ is set
equal to 1, while the interaction constant $\alpha$ has been fixed equal to 20
so as to induce an almost rigid interaction between neighbouring
monomers~\cite{note1}.

The second term expresses the energy cost of local bending; it is described by
the three-body interaction term
\begin{equation}
V_2(\theta_i) = \frac{1 - \cos \theta_i}{16}
\label{v2}
\end{equation}
where  $\theta_i $ is the angle formed between the links connecting
the $(i-1)$--st, the $i$--th and the $(i+1)$--st monomers. In particular, 
\begin{equation}
\cos \theta_i = \frac{ (x_i -x_{i-1})(x_{i+1}-x_i) +
(y_i -y_{i-1})(y_{i+1}-y_i)}{r_{i,i-1}r_{i+1,i}} ,
\label{coste}
\end{equation}
where $-\pi < \theta_i < \pi$.

Finally, heterogeneity is ensured by Lennard-Jones type interaction between
non-neighbouring monomers ($|i-j| >1$)
\begin{equation}
V_3(r_{i,j}) = \frac{1}{r_{i,j}^{12}} - \frac{c_{i,j}}{r_{i,j}^6}
\label{v3}
\end{equation}
where
\begin{displaymath}
c_{i,j} = \frac{1}{2} (2-3\xi_i - 3\xi_j +5 \xi_i \xi_j) 
\quad   .
\end{displaymath}
and $\xi_i = 0$ indicates that the $i$th monomer is hydrophobic (H), while
$\xi_i=1$ corresponds to a polar (P) one. As a result, the interaction is
attractive if the two monomers are either both hydrophobic or both polar
($c_{i,j} = 1$ and $1/2$, respectively), while it is repulsive, if the monomers
belong to different species (in which case $c_{ij} = -1/2$). This potential
choice simulates the effective interaction among H and P monomers in the
presence of a solvent. In fact, since H monomers prefer to avoid a direct
contact with the solvent, they tend to clusterize in the interior where they
can be shielded from water by a shell of P monomers. The net result is an
effective H-H attraction and an H-P repulsion as assumed in the model.

Altogeher, the heteropolymer Hamiltonian writes as
\begin{equation}
H = \sum_{i=1}^L \frac{p_{x,i}^2+p_{y,i}^2}{2m} +
\sum_{i=1}^{L-1} V_1(r_{i,i+1}) + \sum_{i=2}^{L-1} V_2(\theta_i)
+ \sum_{i=1}^{L-2} \sum_{j=i+2}^{L}  V_3(r_{ij},\xi_i,\xi_j)
\label{hamil}
\end{equation}
where all monomers are assumed to have the same mass $m$ and momenta are defined
as $(p_{x,i},p_{y,i}) := m({\dot x}_i,{\dot y}_i)$.

Accordingly, each heteropolymer is perfectly identified by a binary sequence of
0s and 1s specifying the nature of each monomer. Those sequences for which
the heteropolymer shape converges systematically (at intermediate 
temperatures) towards
the same ``native'' configuration independently of the initial condition are
identified as ``good folders''. Previous studies, mostly based on Monte Carlo
techniques indicate that this happens only in a few cases
\cite{irback1,irback2,marinari} and the scenario has been confirmed also by
molecular dynamics simulations \cite{tlp}.

In what follows, we shall limit our investigations to the three following
cases, all of length $L=20$,

\begin{itemize}

\item{[S0]= [0000 0000 0000 0000 0000]} a homopolymer composed of hydrophobic
residues only;

\item{[S1]=[0001 0001 0001 1001 1000]} a sequence first studied in
\cite{irback2} (therein indicated with the code number 81) where it was 
identified as a good folder; 

\item{[S4]=[1110 0100 0000 0001 0010]} a randomly generated sequence with 6
P-type residues, identified as a bad folder in \cite{tlp}.

\end{itemize}

A reasonably accurate characterization of each sequence can be obtained by
determining three transition temperatures. The first one, $T_\theta$, denotes
the temperature below which the polymer is in a collapsed rather than in a
random-coil configuration \cite{degennes}. It can be determined by studying the
temperature dependence of the gyration radius $R_{gy}(T)$: $T_\theta$
corresponds to the maximum of $\partial R_{gy}(T)/\partial T$.

The folding temperature $T_f$ is the temperature below which the heteropolymer
stays predominantly in the native valley. Here, analogously to \cite{tlp}, we
define the native valley as the basins of attraction of the NC and of its
neighbouring minima. A quantitative estimate of $T_f$ can be then obtained by 
determining the temperature at which the chain spends 50\% of the time within
the native valley. 

Finally, the glass-transition temperature $T_g$ can be identified by comparing 
(finite) time averages performed starting from different initial conditions.
Specifically, we have considered unfolding (USs) and folding (FSs) simulations,
whose initial conditions correspond to the NC and to random coil
configurations, respectively.
In practice, $T_g$ is defined as the temperature below which the
relative difference between USs and FSs averages of the internal energy $U$ is
larger than 10\%. 

We have determined $T_\theta$, $T_f$, and $T_g$, by means of molecular-dynamics
simulations with each monomer being in contact with a stochastic thermal
reservoir at temperature $T$ 
\begin{equation}
{\dot z}_i(t) = p_{z,i}/m \quad ; \quad {\dot p}_{z,i}(t) = 
- \frac{\partial H}{\partial z_i} 
- \gamma p_{z,i}(t) + \eta_{z,i}(t)
\label{hj}
\end{equation}
Here $z_i$ is introduced as a shorthand notation for both 
the spatial coordinates $x_i$ and $y_i$,
$\gamma$ is the dissipation rate  and $\eta_{z,i}(t)$ is a Gaussian
distributed,  $\delta$-correlated random noise
\begin{equation}
\label{fd} 
\langle \eta_{z,i}(t)\eta_{z,j}(0) \rangle = 2\gamma mk_BT \delta(t)
\delta_{i,j} ,
\end{equation}
where, $k_B$ denotes the Boltzmann constant and $T$ is the temperature
(for the sake of simplicity, both $k_B$ and $m$ have been set to unit).

\begin{figure}
\cl{
\includegraphics[clip,width=7.truecm]{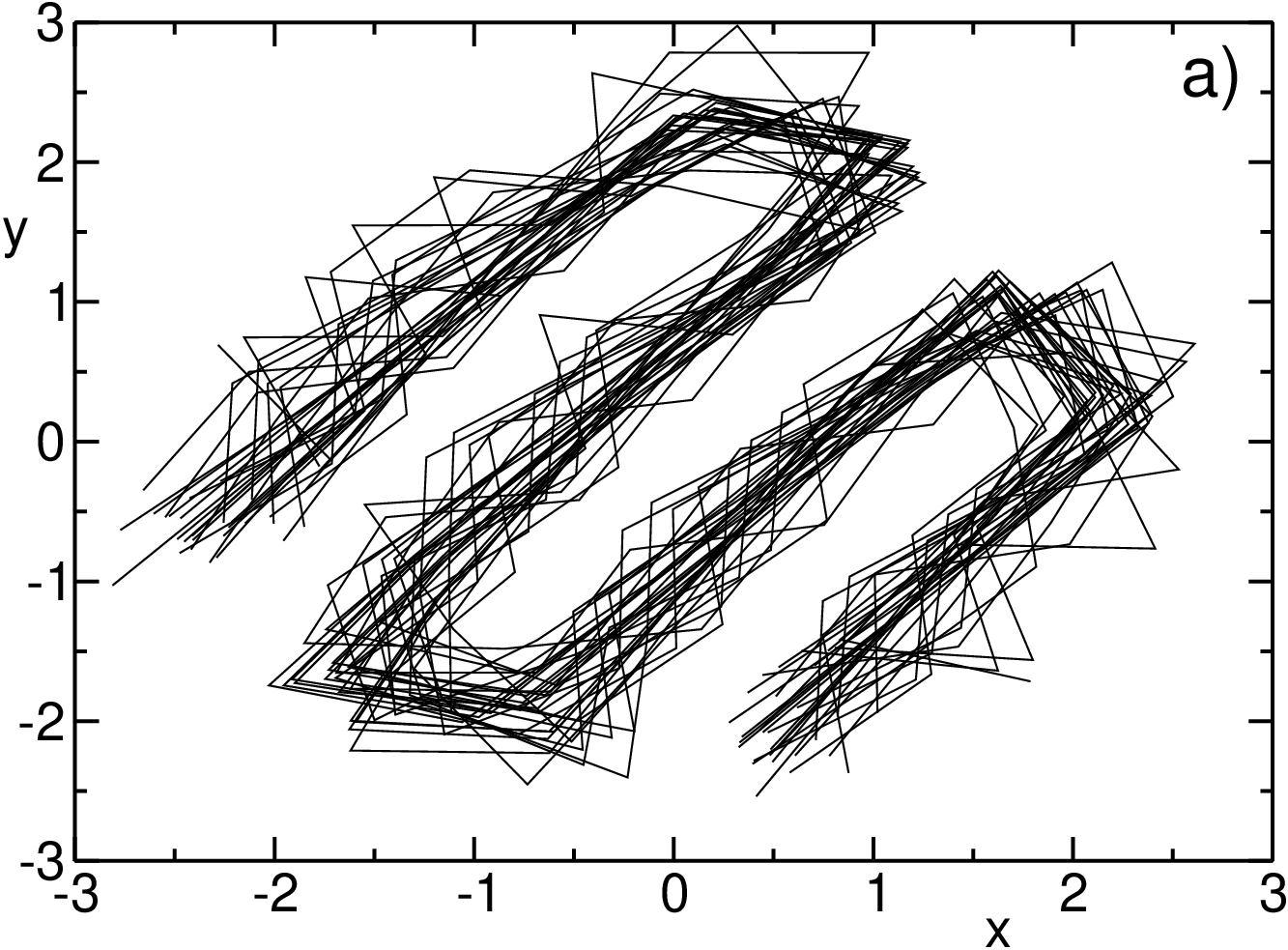}
\includegraphics[clip,width=7.truecm]{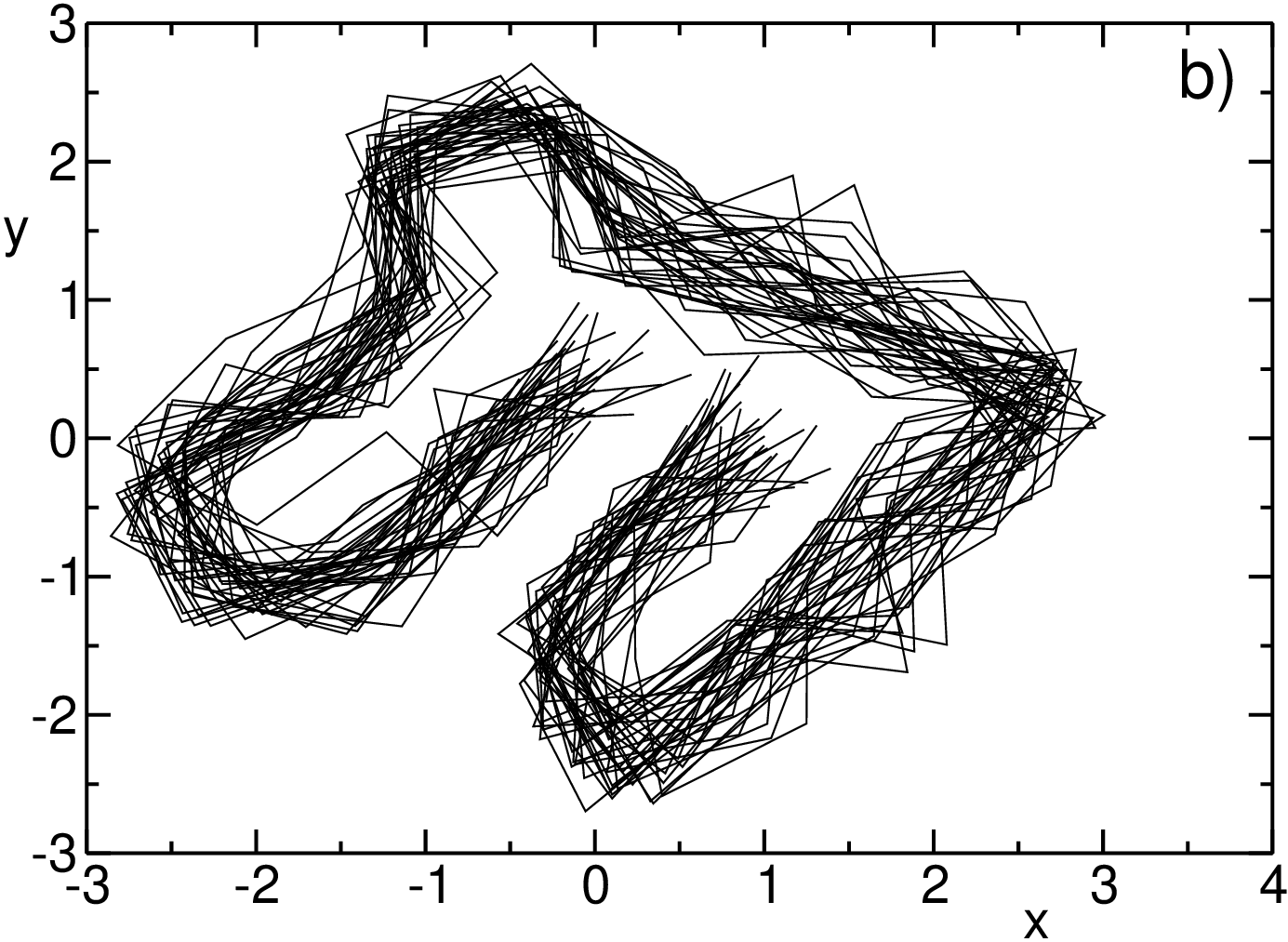}
}
\cl{
\includegraphics[clip,width=7.truecm]{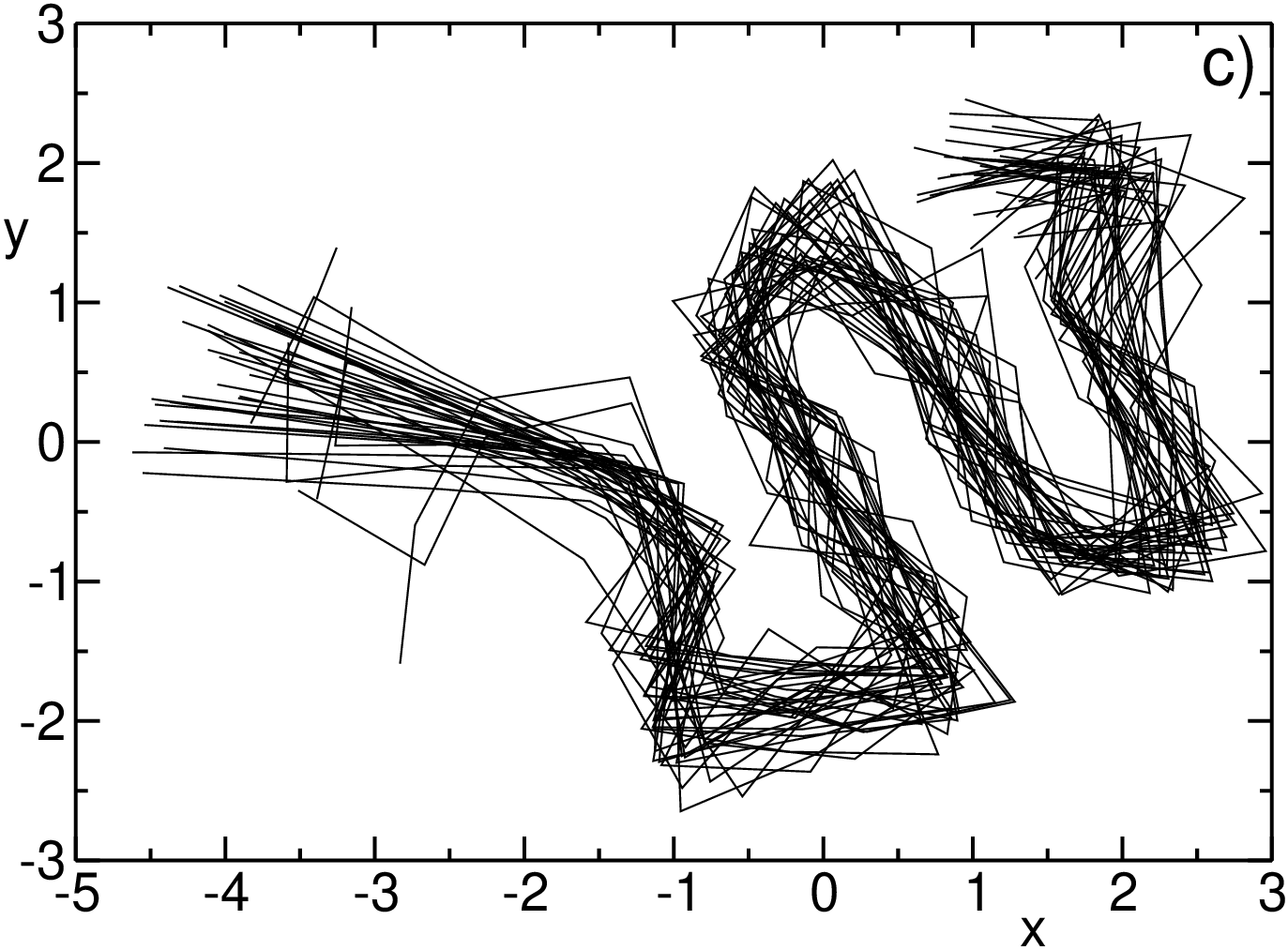}
}
\vskip 0.5 cm
\caption{Configurations of all the NNMs for the three considered
sequences : S0 (a), S1 (b) and S4 (c). }
\label{minima}
\end{figure}

The temperature values obtained for $S0-S1-S4$ are reported in Tab.~\ref{tab1},
together with the number $n_0$ of the minima directly connected with the NC.
The glassy transition has been
determined by performing averages over a time lapse on the order of $10^6$
units. These results are very close to those reported in \cite{tlp}, where a
deterministic Nos\'e-Hoover thermostatting scheme \cite{Nose} was used instead.
The advantage of using the Langevin equation (\ref{hj}) is that the damping rate
can be directly controlled. As we shall see in the next section, this is a
crucial ingredient for characterizing the escape rate from a given valley.
 
\begin{table}[ht]
\label{table}
\caption[tabone]{
The collapse-transition temperature $T_{\theta}$, the glassy temperature
$T_g$, the folding temperature $T_f$, the number $n_0$ of neaerst neighbour
minima of the NC for the sequences S0 (homopolymer), S1 (good folder) and
S4 (bad folder)}
\vskip 0.3 truecm
\begin{tabular}{|c|c|c|c|c|c|}
\hline
 \hfil & \hfil S0 \hfil & \hfil S1 \hfil &
\hfil S4 \hfil \hfil \\
\hline\hline
$T_\theta$ & 0.16  & 0.11 & 0.13 \\
$T_g$ & 0.022  & 0.048 & 0.025 \\
$T_f$ & 0.044  & 0.061 & 0.044 \\
$n_0$ & 31 & 37 &  36 \\
\hline
\end{tabular}
\label{tab1}
\end{table}

\begin{figure}
\cl{
\includegraphics[clip,width=10.truecm]{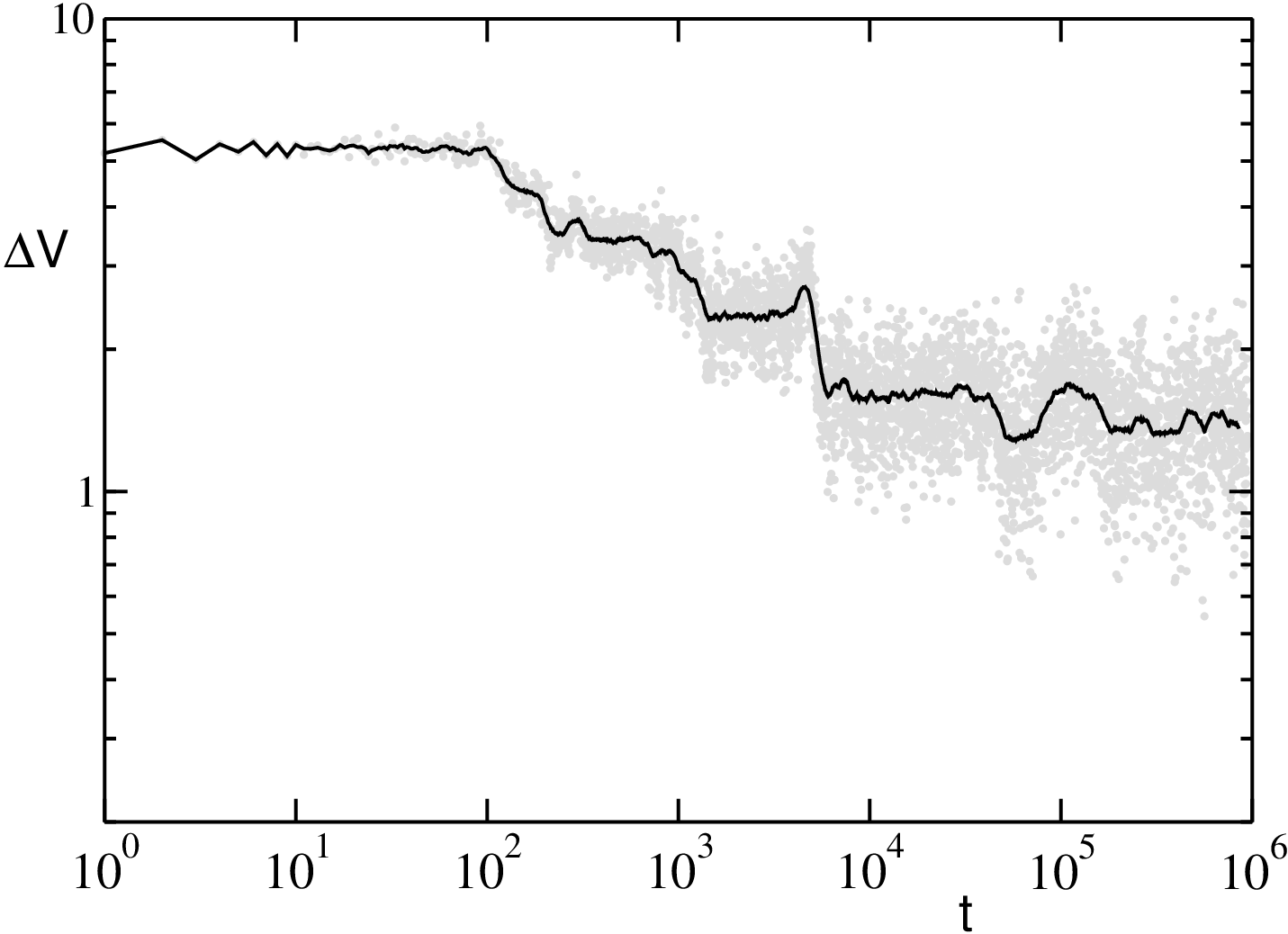}
}
\vskip 0.5 cm
\caption{Potential energy versus time during a folding simulation for
the sequence S1 at $T=0.055$ (grey circles). The solid line is a local average
with exponentially increasing window size.}
\label{pot-81-055}
\end{figure}

A more detailed characterization of heteropolymer dynamics can be obtained by
identifying at least the most visited minima of the potential energy
$V= V_1 + V_2 + V_3$. Here, we have proceeded by sampling a generic trajectory
at time intervals of length $\Delta t \sim  1-5$. Then, the resulting 
instantaneous configurations have been taken as initial conditions for the 
overdamped dynamics
\begin{equation}
{\dot z}_i = - \frac{\partial H}{\partial z_i}  \quad 
\quad , 
\label{overdamp}
\end{equation}
which drives the system to the minimum energy state, whose basin of attraction
contains the initial condition~\cite{still2}. The minima separated from the
absolute minimum (the NC) by a single energy barrier have been denoted as
nearest neighbouring minima (NNM), while those separated from the NC by two
barriers as second-nearest neighbouring minima (2NNM), and so on.
The NNM configurations for all the three examined sequences are reported
in Fig.~\ref{minima}.
Before passing to a specific discussion of the escape from a given valley, it
is convenient to illustrate the outcome of a typical FS in the temperature
range $T_g < T < T_\theta$. The evolution of the difference $\Delta V$ between
the instantaneous potential energy and the potential energy
$V_0$ of the NC is reported in Fig.~\ref{pot-81-055} for the heteropolymer S1.
A series of sudden conformational changes are clearly identifiable from the
various energy drops (notice the logarithmic scale of both axes). Once S1
enters the native valley, it remains there for a very long time, although
jumps towards neighbouring minima can occasionally occur.

\section{Escape rate from a metastable state}
\label{escara}

Since the publication of the pioneering paper of Kramers~\cite{kramers}, the
problem of determining the escape rate from a metastable state has been
addressed in many different contexts. Here, we consider the overdamped
dynamics of an ensemble of $N$ interacting particles in a 2d environment.
We restrict our discussion to the overdamped limit, since it is expected that
in the protein folding problem, the time scale of energy exchanges with
the thermal bath (through collisions with water molecules) is rather fast
(we shall anyway return to this point later on). The
probability density $P({\bf r},t)$ for a configuration to be in
an infinitesimal volume around the state 
${\bf r} \equiv (r_1,\ldots,r_{2N}) = (x_1,y_1,x_2,\ldots,x_N,y_N)$
at time $t$ satisfies the Fokker-Planck equation
\begin{equation}
\frac{\partial P}{\partial t} = \frac{1}{\gamma m}\sum_{i=1}^{2N} \frac{\partial}
{\partial r_i} \left( \frac{\partial V({\bf r})}{\partial r_i} P
+ k_B T \frac{\partial P}{\partial r_i} \right)  \quad ,
\label{fpe}
\end{equation}
where $V({\bf r})$ represents the energy potential. The above is nothing but a
continuity equation, with the r.h.s. representing the divergence of the
probability flux
\begin{equation}
J_i \equiv - \frac{1}{\gamma m} \left[ \frac{\partial V({\bf r})}{\partial r_i}  +
  k_B T \frac{\partial}{\partial r_i} \right] P({\bf r},t) =
-\frac{k_B T}{\gamma m}\exp\left(\frac{-V({\bf r})}{k_B T}\right) \frac{\partial}
{\partial r_i} 
\left[ \exp \left( \frac{V({\bf r})}{k_B T}\right) P({\bf r},t) \right].
\label{cur}
\end{equation}
The stationary solution (with no flux boundary conditions, $J_i=0$) is simply
given by $P({\bf r}) = \exp[V({\bf r})/(k_BT)]$. Let us now assume that the
energy landscape exhibits at least two local minima $m_a$ and $m_b$ with
energies respectively equal to $V_a$ and $V_b$: we want to estimate the escape
rate from the basin of attraction of $m_a$. The boundary separating the basins
of attractions of the two minima coincides with the stable manifold of possibly
more than one saddle point. Let us denote the energy on the saddle with
${\bf r}_s$ is $V_s$. If the system is prepared into the state $m_a$, a flux
$\bf J$ sets in: if the flux itself is weak, it is basically constant in time
and one can approach the problem by determining the stationary state with
$\bf J$ being a solenoidal field. Once $\bf J$ is given, the escape rate $\Gamma$
can be obtained by integrating the outgoing flux over the whole boundary of the
well. We are not aware of a general solution in more than one dimension.

The most general case amenable to an analytic treatment is that of a potential
which, in the vicinity of the saddle point, can be separated into two distinct
contributions
\begin{equation}
V({\bf r}) - V_s = V_\parallel(r_\parallel) + V_\perp({\bf r}_\perp) \quad,
\label{separa}
\end{equation}
where $r_\parallel$ is the distance from the basin boundary (measured along the
unstable manifold of the saddle), while the vector ${\bf r}_\perp$ parametrizes
all other directions in phase space. Finally, the zeroes of $V_\parallel$ and
$V_\perp$ are set in the saddle point. Under the above assumptions, the only
nonzero component of the flux is 
\begin{equation}
J_\parallel = -\frac{k_B T}{\gamma m}\exp\left(\frac{-V({\bf r})}{k_B T}\right)
 \frac{\partial Q}{\partial r_\parallel}
\label{flupe}
\end{equation}
where we have introduced $Q({\bf r}) = \exp(V/k_BT) P({\bf r})$ and $J_\parallel$
depends only on ${\bf r_\parallel}$. The vanishing of $J_\perp$ implies that
$Q$ depends only on $r_\parallel$. Accordingly, Eq.~(\ref{flupe}) can be solved to
yield
\begin{equation}
Q(r_\parallel) = \frac{1}{C} \int_{r_\parallel}^{r_s} {\rm e}^\frac{V_\parallel(\xi)}
{k_BT}d \xi
\label{flunor}
\end{equation}
where the integration constant is determined by imposing that $Q({\bf r})$ and,
accordingly, $P({\bf r})$ vanish along the boundary. The multiplicative 
constant $C$ can be finally determined by normalizing the integral of 
$P({\bf r})$:
\begin{equation}
C = {\rm e}^{-\frac{V_a}{k_BT}} \int {\rm e}^{-\frac{V({\bf r})-V_a}{k_BT}}
d{\bf r} \int_{r_\parallel}^{r_s} {\rm e}^{\frac{V_\parallel(\xi)}{k_BT}} d \xi
\label{ctot} \quad \quad .
\end{equation}
The first integral is restricted to the basin of attraction of $m_a$. In
the small temperature limit, in the region where
${\rm e}^{-\frac{V({\bf r})-V_a}{k_BT}}$ is significantly different from 0, the
last integral is basically constant, so that we can replace its lower border
with $r_a$, thus writing
\begin{equation}
C = {\rm e}^{-\frac{V_a}{k_BT}} I_a I_\parallel
\end{equation}
where
\begin{eqnarray}
I_a &=& \int {\rm e}^{-\frac{V({\bf r})-V_a}{k_BT}} d{\bf r} \\
I_\parallel &=& \int_{r_1}^{r_s} {\rm e}^{\frac{V_\parallel(\xi)}{k_BT}} d \xi
\end{eqnarray}
The flux then writes
\begin{equation}
J_\parallel = \frac{k_B T}{\gamma m}\frac{{\rm e}^{-\frac{\Delta V}{k_BT}}}
 {I_a I_\parallel} {\rm e}^{-\frac{V_\perp}{k_BT}}
\label{flupefin}
\end{equation}
where $\Delta V = V_s - V_a$. By integrating on the basin boundary, one finally
obtains
\begin{equation}
\Gamma = \frac{k_B T}{\gamma m}\frac{I_\perp}{I_a I_\parallel}
{\rm e}^{-\frac{\Delta V}{k_BT}}
\label{erate}
\end{equation}
where
\begin{equation}
I_\perp = \int {\rm e}^{-\frac{V_\perp({\bf r}_\perp)}{k_BT}}
d{\bf r}_\perp
\end{equation}
Eq.~(\ref{erate}) holds under the assumption of a sufficiently small temperature
($\Delta V > k_BT$) and the separability condition (\ref{separa}). The former
hypothesis is needed to ensure a sufficiently slow flux to guarantee that
a quasi-stationary approach holds and the integral in Eq.~(\ref{ctot})
factorizes. The latter hypothesis ensures that Eq.~(\ref{flunor}) indeed
represents a meaningful solution of the problem. One should notice that
separability is required to hold only in the region around the saddle where
the integral $I_\perp$ has to be performed.

If the temperature is small enough, only the leading quadratic terms are
relevant in the computation of the integrals $I_a$, $I_\parallel$ and $I_\perp$.
In this harmonic limit, they reduce to Gaussian integrals that can
be computed by diagonalizing the Hamiltonian. Upon denoting with
$\omega_a^{(i)}$ the $2N$ frequencies in the vicinity of the minimum
$m_a$ ($[\omega_a^{(i)}]^2 = -\Lambda_a^{(i)}/m$, $\Lambda_a^{(i)}$ being the
negative $i$th eigenvalue of the Hessian), with $\omega_\perp^{(i)}$ the $2N-1$
frequencies around the saddle (those corresponding to the stable directions),
and with $\omega_\parallel$ the rate associated to the only expanding direction
($[\omega_\parallel]^2 = \Lambda_s^{(2N)}/m$), one obtains the expression
first derived by Langer in 1969~\cite{langer,langer1} \cite{note2}.
\begin{equation}
\Gamma_L = \frac{\omega_\parallel}{\pi \gamma} 
\frac{\prod_{i=1}^{2N} \omega_a^{(i)}}{\prod_{i=1}^{2N-1} \omega_\perp^{(i)}}
\rm{e}^{-\frac{\Delta V}{k_B T}} \equiv \frac{\omega_\parallel^2}{\pi \gamma R}
\rm{e}^{-\frac{\Delta V}{k_B T}}
\quad ,
\label{rate_n}
\end{equation} 
where $R:=\sqrt{\prod_{i=1}^{2N}|\Lambda^{(i)}_s|/
\prod_{i=1}^{2N}|\Lambda^{(i)}_a|}$, can be interpreted as an entropy factor
\cite{parisi,ruocco}. In the case of continuous symmetries (as, e.g.,
translational and rotational symmetries), Hessian eigenvalues corresponding to
Goldstone modes vanish. They have to be excluded in the frequency products
appearing in Eq.~(\ref{rate_n})~\cite{langer1}.

Expression (\ref{rate_n}) is routinely employed in studies of many-body
systems, including relaxation dynamics in glasses \cite{parisi} and in the
estimation of entropy barriers in clusters \cite{ruocco}. However, its validity
range has not been thoroughly investigated. For instance, in Ref.~\cite{berry}
a Master equation is constructed for a cluster of 19 atoms by identifying
directly minima and saddles, but estimating transition probabilities
only from an expression similar to Eq.~(\ref{rate_n}).

In the weak damping limit, Eq.~(\ref{rate_n}) generalizes to \cite{review}
\begin{equation}
\Gamma_L = \frac{\zeta \omega_\parallel^2}{\pi \gamma R}
   \rm{e}^{-\frac{\Delta V}{k_B T}}
\label{langernew}
\quad ,
\end{equation}
where the multiplicative correction 
\begin{equation}
\zeta = \frac{2}{1 + \sqrt{1 + (2\omega_\parallel/\gamma)^2}} ,
\label{weakda}
\end{equation}
depends on the ratio between the damping constant and the divergence rate along
the expanding direction.

Finally, we mention a simplified formula for the escape rate that is somehow
halfway between the general expression (\ref{erate}) and that one
corresponding to the linearization (\ref{rate_n}). Under the hypothesis of a
fully separable potential, one can factorize the two integrals $I_a$ and
$I_\parallel$ into products of one-dimensional integrals
\begin{eqnarray}
I_a &=& \prod_{i=1}^{2N} \int {\rm e}^{-\frac{V_i(r_i)}{k_BT}} d r_i 
\label{rate_1da}
\\
I_\parallel &=& \prod_{i=1}^{2N-1} \int
 {\rm e}^{-\frac{V^{(i)}\left(r^{(i)}\right)}{k_BT}} d r^{(i)}
\label{rate_1db}
\end{eqnarray}
where $V_i$ is the $i$th component of the potential in the vicinity of the
minimum with the zero of the scale such that $V_i(0) = 0$, and $V^{(i)}$ is
the analogous around the saddles for the stable directions.

In the following section we compare Langer prediction (\ref{rate_n}) with our
numerical results. In Sect. V we discuss the limits of applicability of
Eq.~(\ref{rate_n}) and we test the improved expression for the escape rate
(\ref{erate}) (with the integrals estimated as in
Eqs.~(\ref{rate_1da},\ref{rate_1db})).

\section{Numerical Results}
In order to characterize the dynamics of the polymer in the native valley
we have determined both analytically and numerically the escape rate from
the NC towards any of the NNMs. The procedure for identifying the NNMs from a 
database of ``inherent'' minima, constructed by following the method outlined
in Sect. II, relies on the identification of the minimal energy path connecting
each NNM to the NC. The algorithm used to find these paths is described in
the Appendix: it allows identifying the saddle separating any two minima.

A few pairs of NNMs turn out to be connected by more than one (up to 3)
minimal energy path, this implying that they are separated by more than one
saddle. In these cases, one should, in principle, compare the numerically
determined escape rate with the sum of the probability flows $\Gamma_L$ through
the different saddles. As a matter of fact, we limited to consider the
contribution of the saddle yielding the maximal flow. This approximation 
is definitely negligible with respect to the discrepancies between
numerical and theoretical estimates fully discussed in the following.

\begin{figure}[ht]
\center{\includegraphics[clip,width=8.truecm]{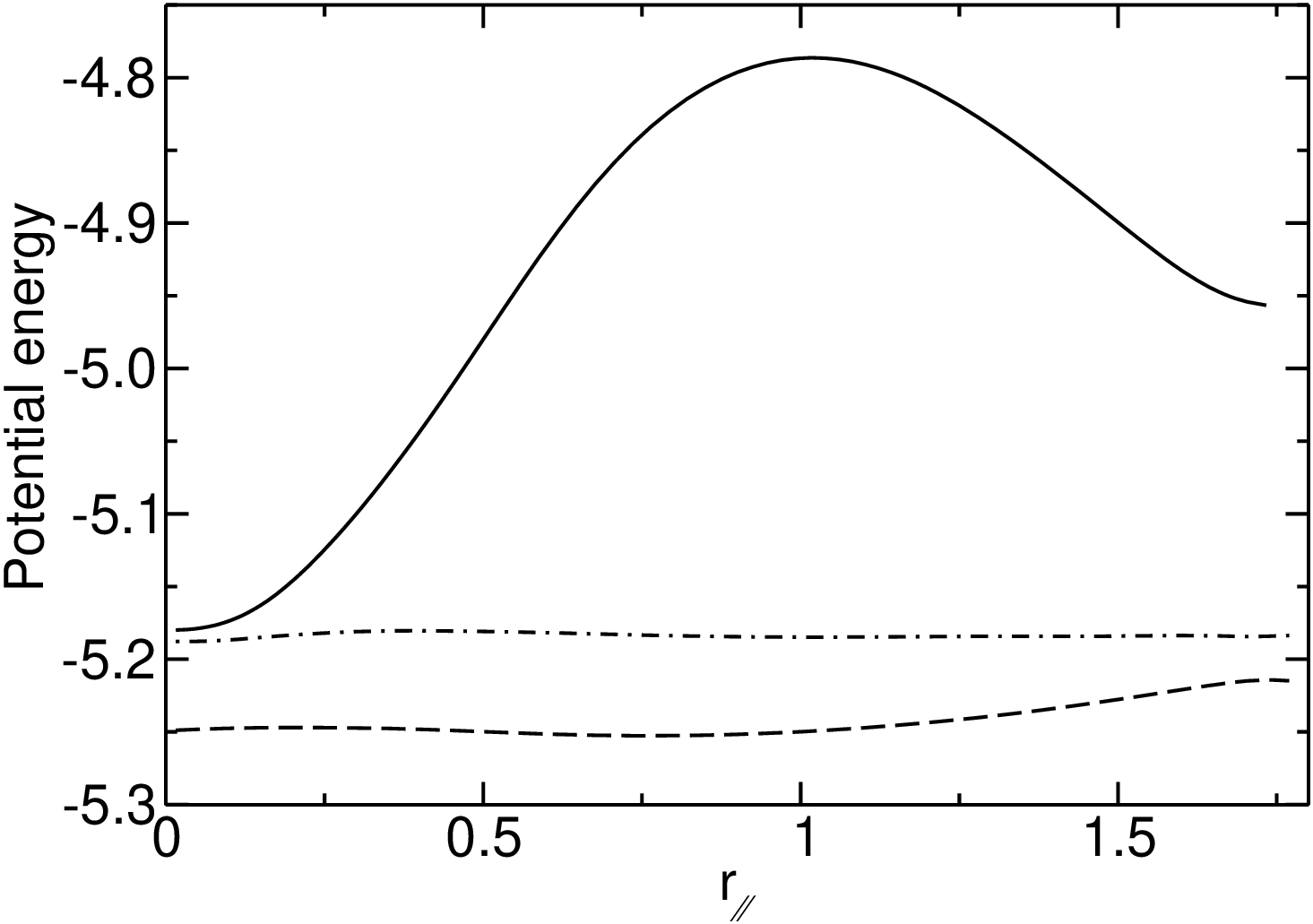}
}
\caption{Potential energy profiles versus
the distance from the NC (measured along the
unstable manifold of the saddle) $r_\parallel$.
The three curves correspond to the different potential
contributions for a minimal energy path connecting the NC
to a nearest neighboring minimum for the sequence S1. The solid
line indicates the Lennard-Jones contribution $V_3$~(\protect\ref{v3}),
while the dashed line the harmonic term $V_1$~(\protect\ref{v1}) and
the dot-dashed line the potential term $V_2$~(\protect\ref{v2}).
The last two terms are shifted along the vertical axis by
a factor $-5.2$ and $-5.75$, respectively.
}
\label{profile}
\end{figure}

In Fig.~\ref{profile} the three potential contributions $V_1$, $V_2$, and $V_3$
are reported for a minimal energy path connecting the NC to one of the NNMs. A
common feature to all the examined paths is that the main contribution to the 
potential energy barrier arises from the Lennard-Jones term. This confirms that
the term driving the folding is indeed the long-range potential term mimicking
the hydrophobicity effects (as already mentioned in \cite{tlp}).

\begin{figure}[ht]
\center{
\includegraphics[clip,width=8.truecm]{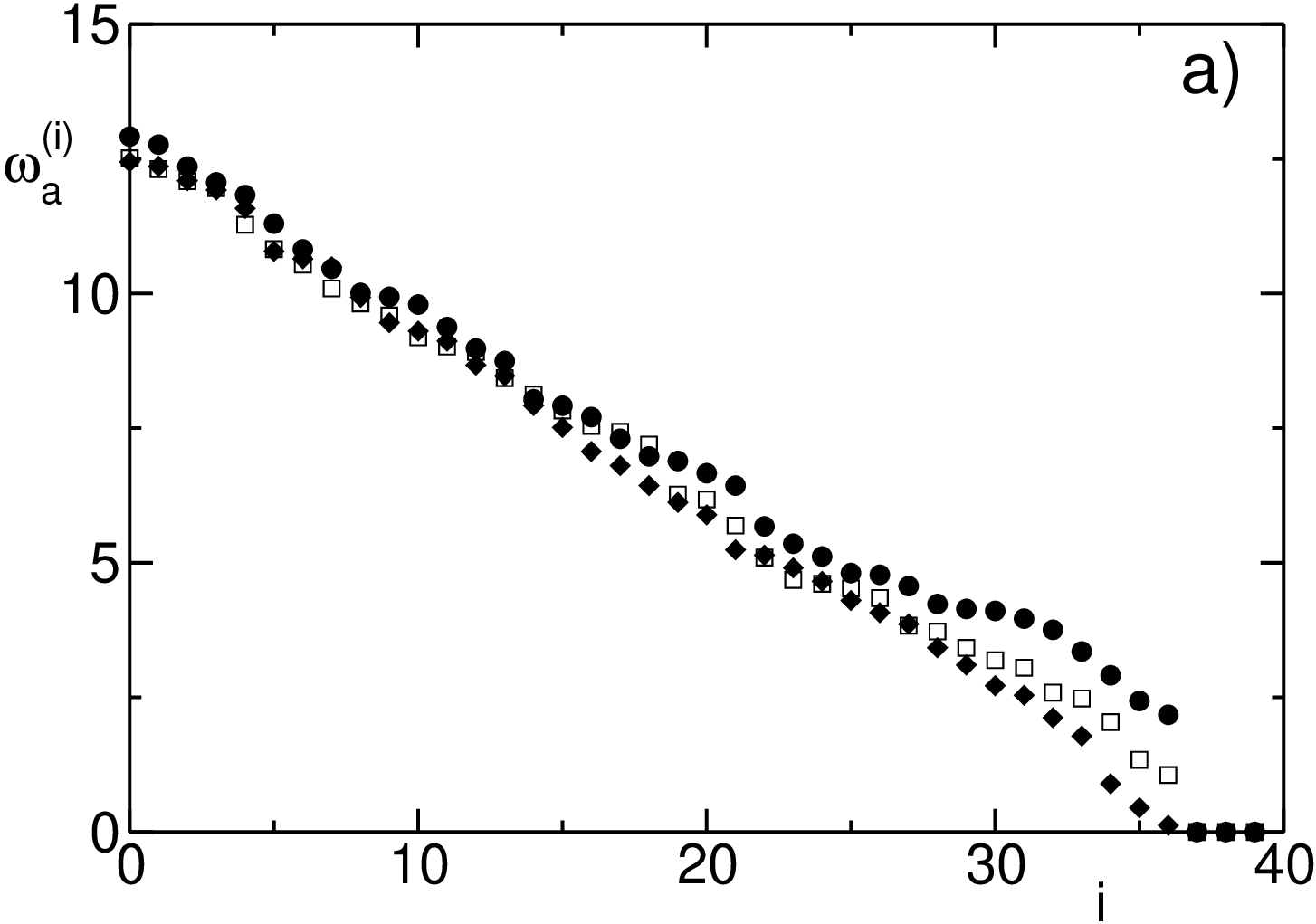}
\includegraphics[clip,width=8.truecm]{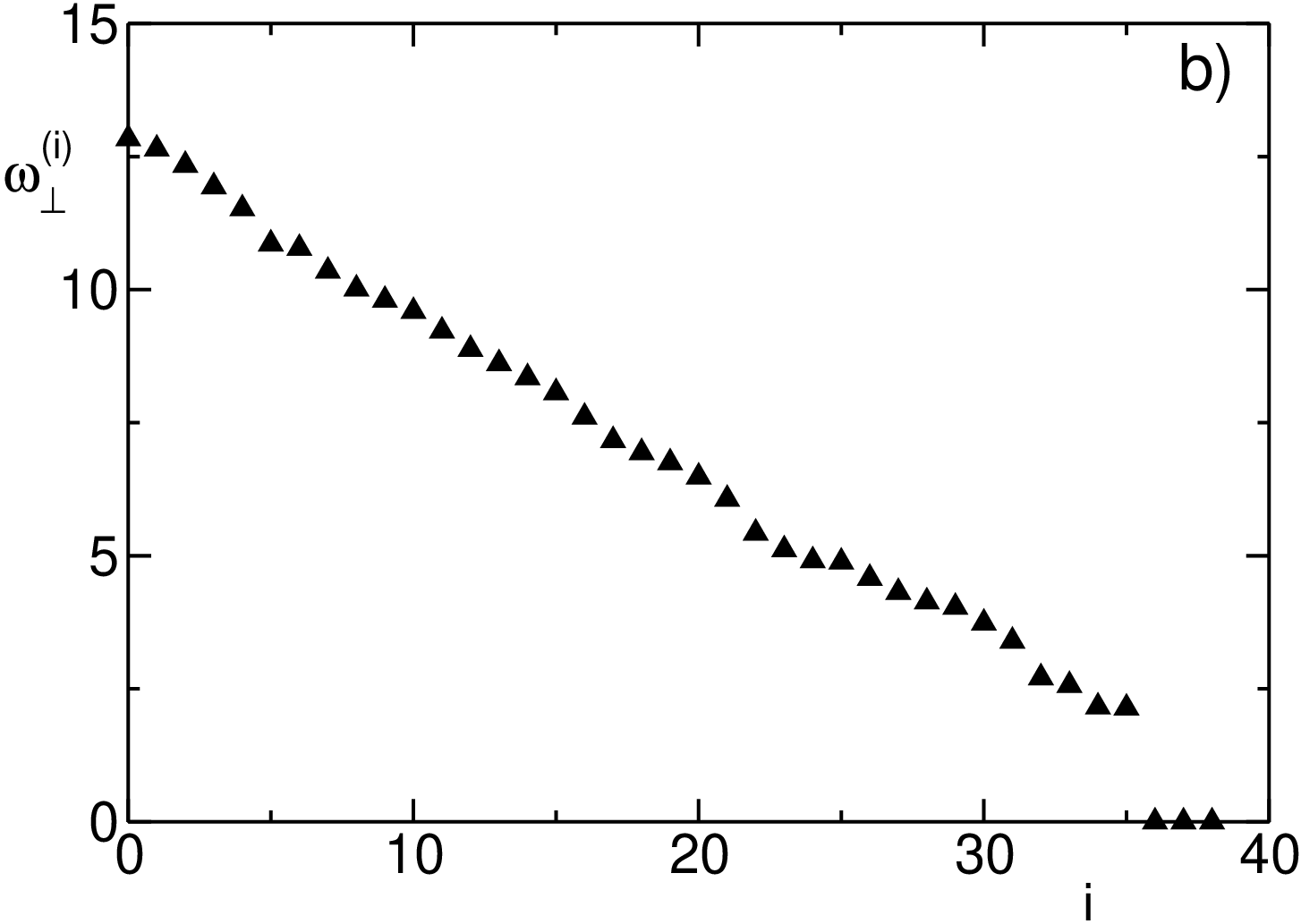}
}
\caption{a) Angular frequencies $\omega_a^{(i)}$ associated to the Hessian of
the NC for the sequence S0 (squares), S1 (circles) and S4(diamonds); b) angular
frequencies
$\omega_\perp^{(i)}$ associated to the Hessian of a saddle separating
the NC from a NNM for the sequence S1.
}
\label{spet}
\end{figure}

The evaluation of $\Gamma_L$ requires the knowledge of the eigenvalues of
the Hessian in both the NC and a suitable saddle. In our model, because of
translational and rotational symmetries, three eigenvalues always vanish. This
is clearly seen in Fig.~\ref{spet}, where the frequency spectrum $\omega_a^{(i)}$
of the NC is plotted for the three sequences (panel a), together with the
spectrum $\omega_\perp^{(i)}$ of one saddle (panel b). The $\omega_a^{(i)}$
spectrum decreases smoothly from values around 13 down to 0. It is interesting
to notice that all spectra do not differ significantly from what one would
obtain for a purely harmonic chain, in which case the spectrum would decrease
from a maximum frequency equal to $2\sqrt{2\alpha} = 12.6$ down to zero.
It is only at lower frequencies that differences among the various sequences
can be appreciated: in fact, this spectral band is basically determined
by the angular motion that is primarily controlled by the cosine (\ref{v2}) and
Lennard-Jones (\ref{v3}) potentials.

For what concerns the value of $\omega_\parallel$, it turns out to range between
$0.3$ and $1.8$ in all saddles.        

Then, we have directly determined the escape rate $\Gamma(j)$ from the
NC by performing several USs with the damping constant set equal to 6.9.
Every $\Delta t$ time units, the ``inherent'' polymer configuration is
determined by a steepest descent method. As soon as the polymer leaves the
basin of attraction of the NC, the corresponding time is registered together
with the new minimum that has been reached. Let us denote with $M_j$ the 
number of
USs ending in the $j$-th minimum, with $\langle \tau_j \rangle$ the
corresponding average escape time and with $ M =\sum_{j=1}^{n_0} M_j $ 
the total number of USs. We have verified that
$\langle \tau_j \rangle$ is independent of $j$. This 
indicates that the polymer spends most of the 
time before any jump in exploring the same region in the phase space,
consistently with the hypothesis of a thermally activated process.
Thus, the escape rate toward the $j$-th minimum can be numerically 
estimated as
\begin{equation}
\Gamma(j) = \frac{M_j}{M}\frac{1}{\langle \tau \rangle} \quad ,
\label{rate_num}   
\end{equation}
where we have dropped the dependence on $j$ in the escape time.

We have focused our numerical analysis in the temperature range
$[T_g,T_\theta]$, where the polymer spends most of the time in a collapsed
state. In order to obtain a sufficient statistics
the parameters of USs have to be suitably tuned according to the
temperature values. In particular, the sampling time $\Delta t$ has
to be maintained sufficiently short to avoid back-crossings of the barrier and
multiple jumps. More precisely, it must be smaller than the lifetime of all
NNMs. Since all lifetimes decrease with the temperature, increasingly smaller
$\Delta t$'s have to be adopted when the temperature is increased~\cite{note3}.
We have chosen $\Delta t$-values ranging
from $1$ at low temperatures ($T \simeq T_g$), to $10^{-3}$ at high
temperatures ($T \simeq T_\theta$), while the integration time step has been
kept equal to $10^{-3}$ (a few tests performed with an integration time step
$\simeq 10^{-4}$  have not revealed any relevant difference).
Anyway, at low temperatures the escape
rate towards a subset of NNM is so small that in practice they are never
found to be visited over an extremely large number of simulations. This is
why in Table ~\ref{tab-times} we indicate the number $n_v$ of visited NNM
as a function of the tempareture: it turns out
that all NNM are visited already at $T=0.08$ for both S0 and S1,
while for S4 new minima are found up to $T=0.1$ (the highest temperature we 
have examined).

\begin{figure}
\cl{
\includegraphics[clip,width=8.truecm]{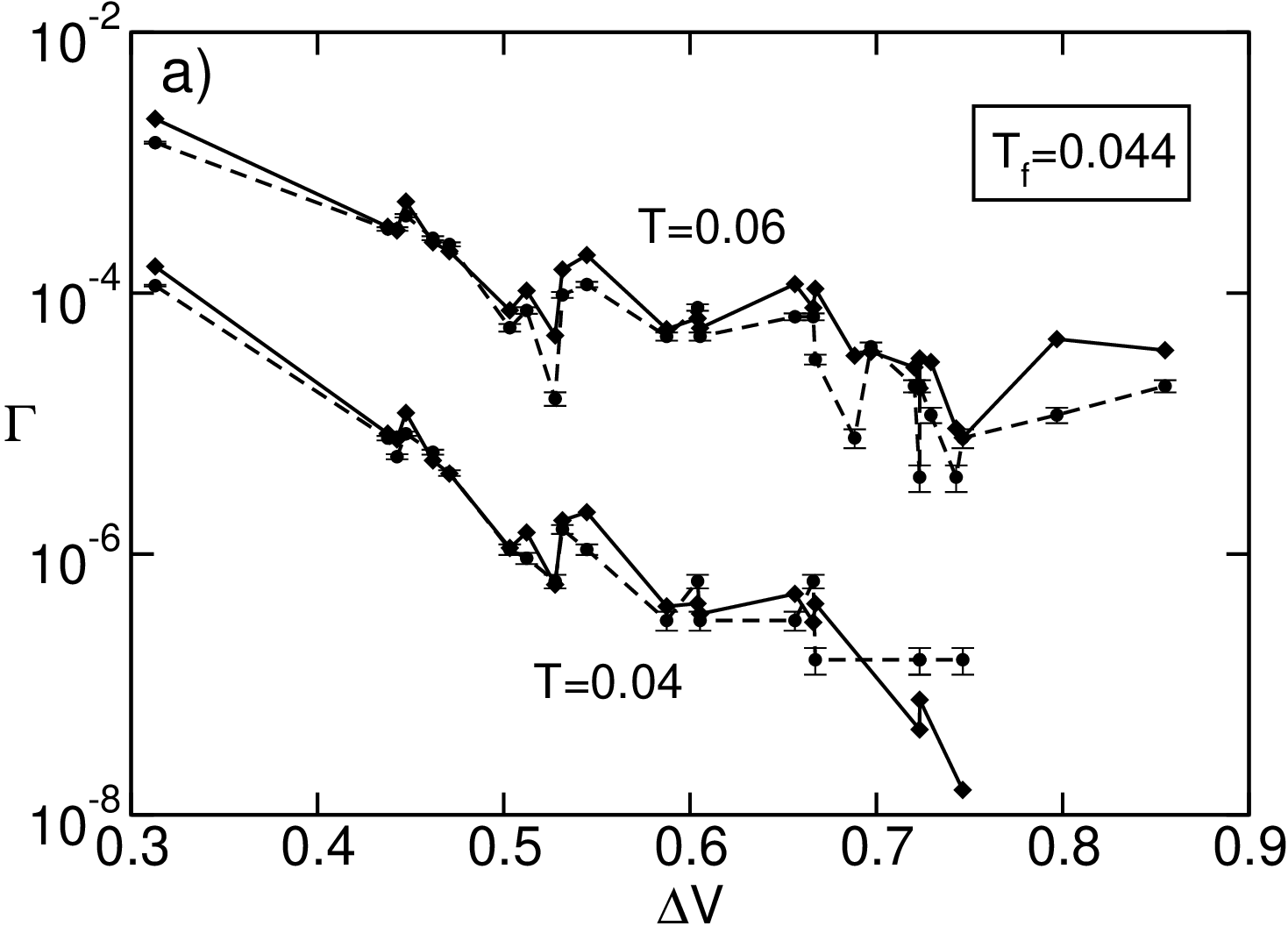}
}
\cl{
\includegraphics[clip,width=8.truecm]{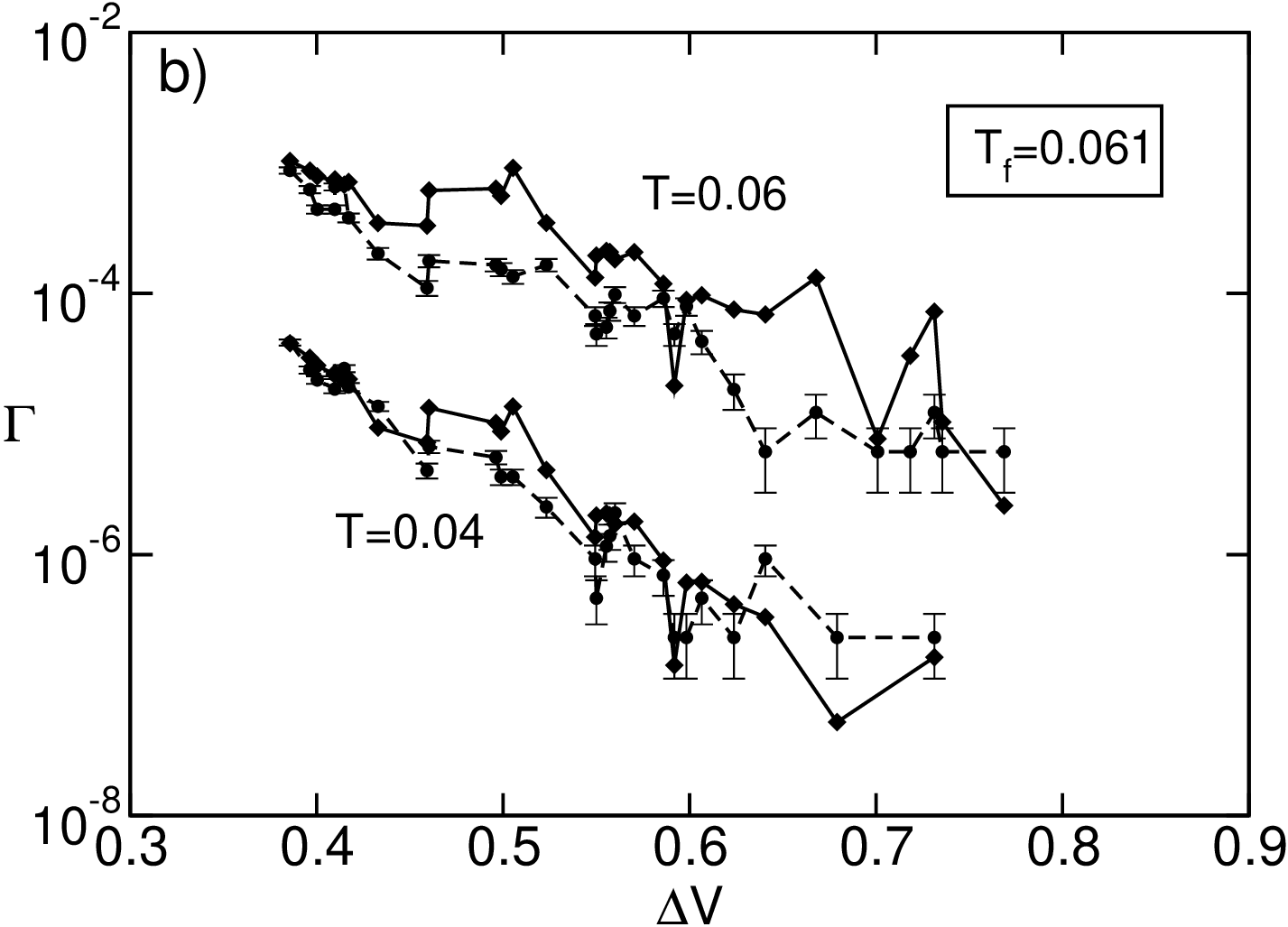}
\includegraphics[clip,width=8.truecm]{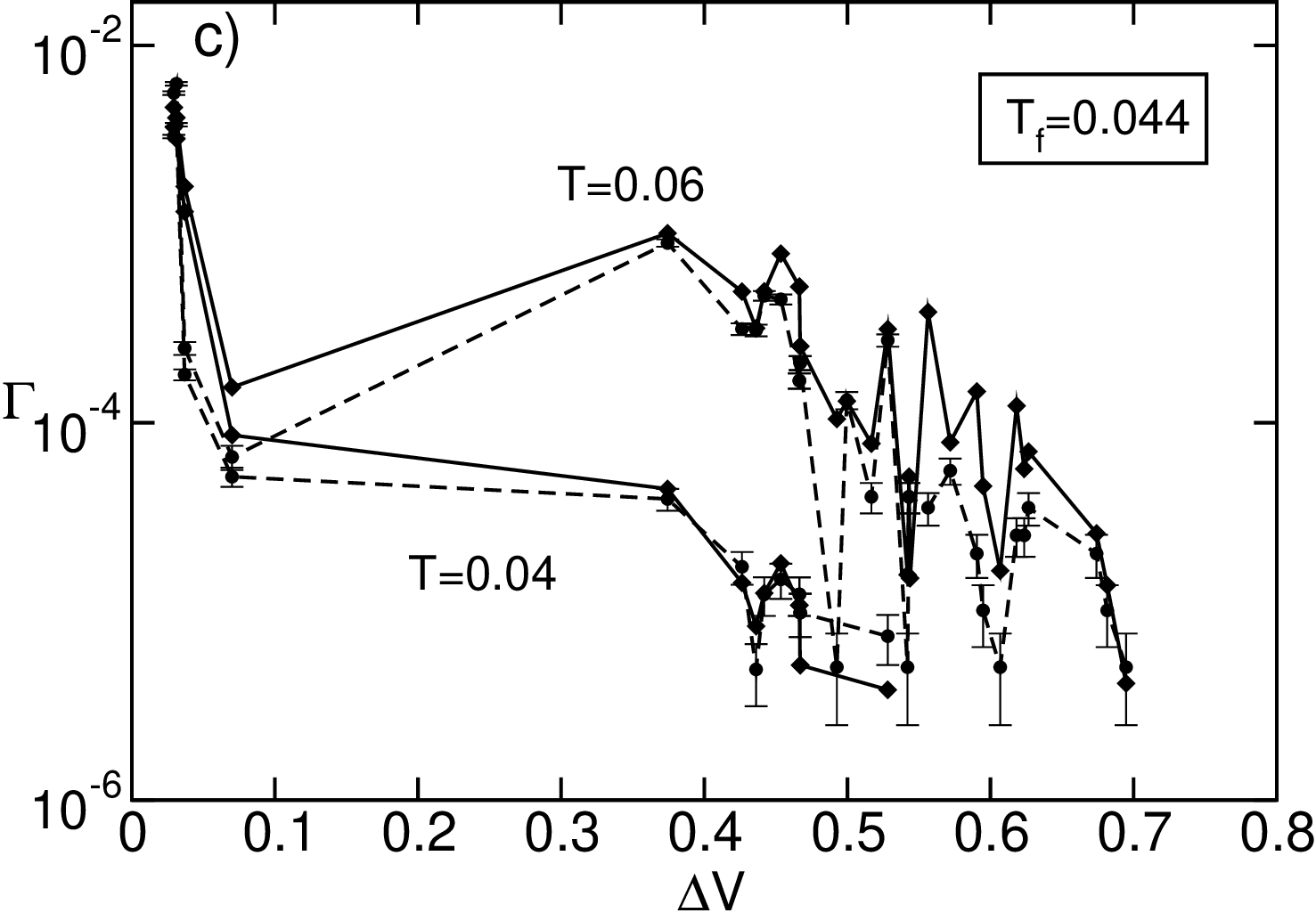}
}
\caption{Transition rates $\Gamma$ versus the barrier height $\Delta V$:
comparison of the theoretical expression
(\ref{rate_n}) (solid line) with the numerical estimate
(\ref{rate_num}) (dashed line) for two
different temperatures ($T=0.04$ and $T=0.06$) for the three
studied sequences. Namely, (a) S0; (b) S1 and (c) S4.
The numerical data refer to a total number $M=1,000 - 5,000$ of
USs and to a sampling time  $\Delta t=1$.
}
\label{prob}
\end{figure}

The numerically computed transition rates are presented in Fig.~\ref{prob}, 
where only the results for the most visited NNMs are reported (see the dashed 
lines). The solid lines refer to the theoretical estimates. We expect
that Eq.~(\ref{rate_n}) holds, since the chosen value of the damping
coefficient, $\gamma=6.9$, should guarantee an overdamped dynamics.

While at $T \simeq 0.04$, the analytic expression $\Gamma_L$ is in good 
agreement with the numerical estimates (for all of the three sequences), at 
higher temperatures the theoretical expression overestimates the escape rate. 
In order to perform a quantitative comparison, it is convenient to compute
the ratio 
\begin{equation}
r_j = \frac{\Gamma_L(j)}{\Gamma(j)}  \quad ,
\label{rel_er}
\end{equation}
for each escape process towards an NNM and to then average over all 
neighbouring minima 
\begin{equation}
\langle r \rangle = \sum_j P_j r_j \quad ,
\end{equation}
where $P_j=\Gamma(j)/\sum_i \Gamma(i)$ is the probability that an US ends up in
the $j$-th NNM and the sum is restricted to those minima that have been visited
at least twice. The values of $\langle r \rangle$ for the three sequences
at four different temperatures from 0.04 up to 0.1 are reported in
Tab.~\ref{tab-times}. In this range, statistically reliable estimates are 
obtained already for $M > 10^3$. In all cases, the theoretical formula is
reasonably accurate ad low temperatures, but it downgrades when going above 
the ``folding'' temperatures and the phenomenon is more evident in the case of
the good folder S1.

\begin{table}[ht]
\caption[tabtwo]{
For all the sequences and four different temperatures $T$ we report: 
the numerically estimated average escape time $\langle\tau \rangle$, the
number of visited NNM $n_v$ (the numbers within brackets refer to the 
minima visited more than once) and the weighted ratio $\langle r \rangle$.}
\vskip 0.3 truecm
\begin{tabular}{|r|r|r|r|r|}
\hline
\hfil $T$ \hfil \hfil & \hfil $\langle\tau \rangle$ \hfil & \hfil $n_v$ \hfil &
\hfil $\langle r \rangle$  \hfil \hfil \\
\hline
\hline
S0 - $0.04$ &  6484 &  20(16) &  1.37 \\
$0.06$  &   265 &  27(25) & 1.42 \\
$0.08$  &    42 &  31(30) &  1.76 \\
$0.10$  &    19 &  31(30) &  3.43 \\
\hline
\hline
S1 - $0.04$ &  4375  &  28(23) &  1.31 \\
$0.06$  &  168  &   32(27) &  1.95 \\
$0.08$  &   24  &   37(34) &  2.05 \\
$0.10$  &   19  &   37(36) &  3.89 \\
\hline
\hline
S4 - $0.04$ &  129    &  12(11) &  1.30 \\
$0.06$  &  58     &  28(24) &  1.42 \\
$0.08$  &  30     & 27(25)  &  2.16 \\
$0.10$  &  18   &  30(24)    &  3.74  \\
\hline
\end{tabular}
\label{tab-times}
\end{table}

\section{Discussion of the Results}

In this section we discuss several independent factors that can a priori affect
the observed escape rates. First of all, we have investigated whether intrinsic
fluctuations due to the chaotic dynamics contribute significantly to the escape
rates. In order to examine this point, we perfromed microcanonical simulations
at various temperatures (namely $T=0.04$, $0.06$, and $0.08$). Although the
potential energy is larger than the barrier height, no jumps have been observed
in simulations lasting up to $2\cdot10^6$ time units. This means that local
fluctuations are not strong enough to trigger jumps between neighbouring
valleys in the presence of a global energy conservation. Therefore,
fluctuations due to the coupling with independent heat baths are a crucial
ingredient in establishibg the time scale of the escape rate.

Yet the observed discrepancy between the analytic expression (\ref{rate_n}) and
the numerically evaluated escape rate calls for an explanation.
Eq.~(\ref{rate_n}) has been derived by making several assumpation that may not
be fulfilled in practice:

\begin{enumerate}
\item the value of the friction coefficient $\gamma$ should be
larger than the frequency associated to the expanding direction
of the saddle $\omega_\parallel$, i.e. the dynamics should be
overdamped;

\item the configurational probability density $P({\bf r},t)$ 
should be almost stationary, i.e. the polymer should be well thermalized
before a jump occurs;

\item the potential energy should be well approximated by the quadratic
contributions in the relevant regions around both the saddles and the NC;
\end{enumerate} 

In the following we investigate the validity of these assumptions.

\subsection{Overdamped Limit}

Throughout this paper we have fixed $\gamma$ equal to 6.9 in adimensional units.
In order to check whether this is a meaningful choice in the protein context,
we must express the damping rate in physical units, $\gamma = 6.9 m/\tau_o$,
where $m$ is the mass of a typical aminoacid, while $\tau_o$ is the period of
small oscillations within the potential well. Since $m \sim 10^{-22}$g and
$\tau_o \sim 10^{-12}$ s, 
it follows that $\gamma \sim 10^{-9}g/s$, a value to be
compared with the typical relaxation rate due to collisions with water
molecules $\gamma_{H_2O} \simeq 10^{-8}-10^{-9} g/s$ (see Ref.~\cite{veit} for
a more detailed discussion). Our choice of $\gamma$ is, therefore, not too far
from reality. Moreover, since $\gamma$ is already four times larger than the
maximum $\omega_\parallel$, we expect the system dynamics to be in the
overdamped regime.

\begin{figure}
\cl{
\includegraphics[clip,width=8.truecm]{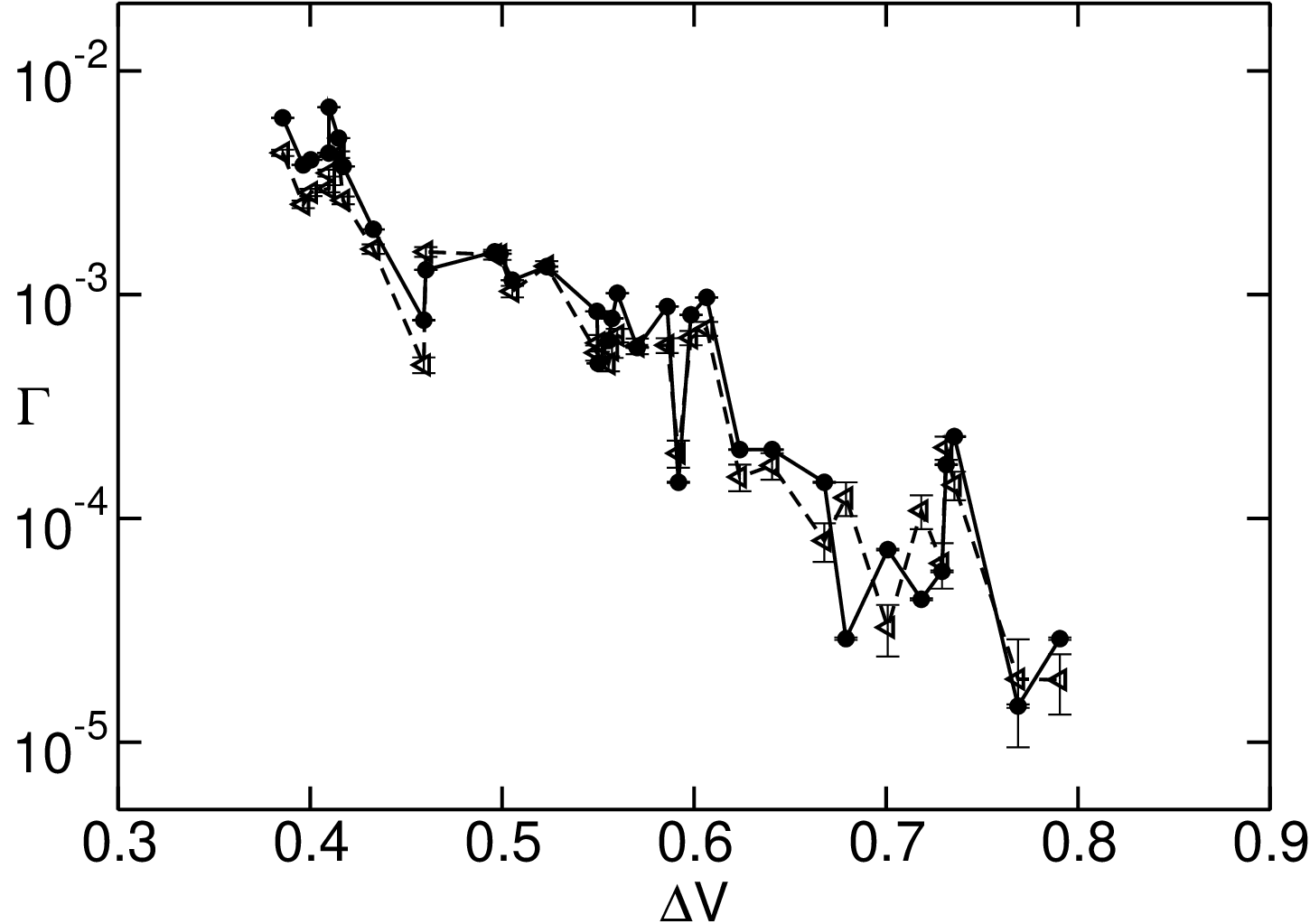}
}
\vskip 0.5 cm
\caption{Numerical transition rates $\Gamma$ vs. the barrier height $\Delta V$
for two different $\gamma$-values. Filled circles refer to $\gamma=1$, while
empty triangles correspond to $\gamma=49$. The latter have been scaled by a
factor $49\times 0.56$ (see text). 
Simulations refer to sequence S1 at $T=0.07$. The
number of USs is $M \simeq 4,000$.}
\label{confrogamma}
\end{figure}

Anyhow, it is instructive to investigate whether the damping rate is responsible
for the non perfect agreement between numerical data and the approximate
theoretical expression. In order to clarify this point we performed further USs
with both a smaller ($\gamma =1$) and a larger ($\gamma=49$) friction. Since
$\gamma$, in the former case, is of the order of $\omega_\parallel$, one should
merely expect multiplicative corrections arising from the $\zeta/\gamma$ factor
in Eq.~(\ref{langernew}). By neglecting saddle-to-saddle fluctuations of
$\omega_\parallel$ (assumed always equal to its average value 1.17), this
amounts to a correction term $49\times 0.56$~.
In Fig.~\ref{confrogamma}, we have multiplied by this factor the numerical data
obtained for $\gamma = 49$. The quite good overlap between the two sets of data
confirms that the dependence on the damping term is well reproduced by the
theoretical formula (part of the oscillations is of statistical nature and
part is due to the neglected $\omega_\parallel$ fluctuations).

\subsection{Thermalization time}

A fundamental hypothesis implicitely made in the derivation of the analytical
expression for the escape rate is that the probability density of initial
conditions inside the native valley can be approximated by the product of the 
Boltzmann-Gibbs distribution times a factor that differs significantly from
1 only in the neighborhood of the saddle. In order to verify the validity of
this assumption, we have studied the decay of the autocorrelation functions
for the kinetic and potential energy in several USs with S0 and S1.

\begin{table}[ht]
\caption[tabcorr]{
Decay times for the autocorrelation function of the kinetic energy $\tau_K$ 
and potential energy $\tau_V$, for S0 and S1 
at various temperatures $T$. The times $\tau_{K,V}$ are estimated by assuming
an exponential decay for the autocorrelation functions.  The functions
has been obtained by averaging over $M=10$ different USs.
}
\vskip 0.3 truecm
\begin{tabular}{|r|r|r|r|r|r|r|r|}
\hline
  T \hfil  &  $\tau_K$ - S0  & $\tau_V$ - S0  &  $\tau_K$ - S1  & $\tau_V$ - S1 \\
\hline
\hline
0.02     &    0.13 &  0.70 & 0.10 & 0.65  \\
0.04     &    0.12 &  1.16 & 0.10 & 0.95  \\
0.06     &    0.14 &  2.17 & 0.10 & 1.55  \\
0.08     &    0.13 &  2.66 & 0.10 & 1.52  \\
\hline
\end{tabular}
\label{tab-corr}
\end{table}

\begin{figure}
\cl{
\includegraphics[clip,width=8.truecm]{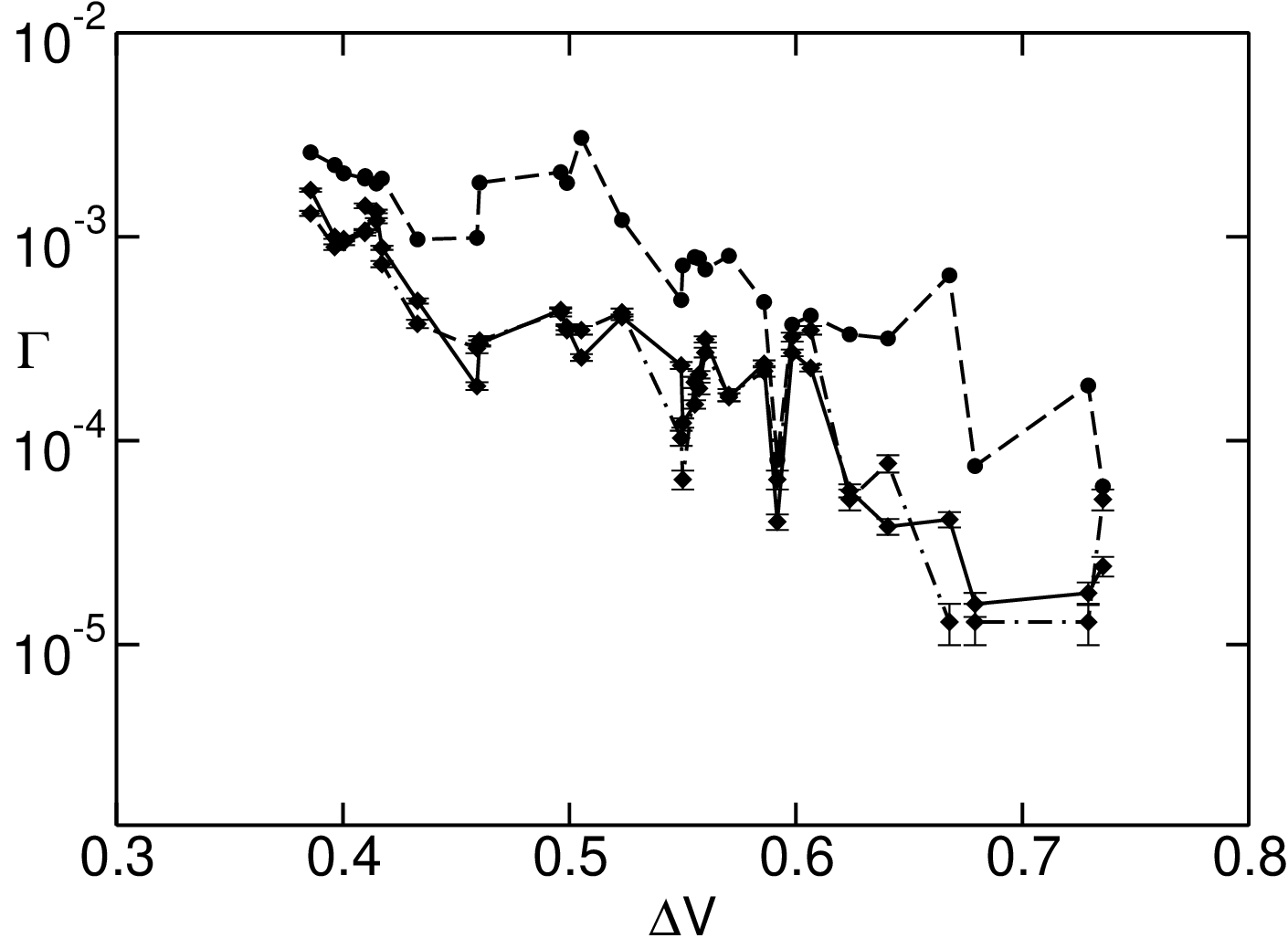}
}
\vskip 0.5 cm
\caption{Transition rates $\Gamma$ versus the barrier height $\Delta V$.
Langers's expression (\ref{rate_n}) (dashed line) is compared with the 
numerical estimate for two choices of initial conditions: with (dot-dashed 
line) and without (continuous line) thermalization. Data refers to the 
sequence S1 at temperature $T=0.07$ with $M = 1,000$ and a thermalization 
time $T= 100$.
}
\label{confro-term}
\end{figure}

From the data reported in Tab.~\ref{tab-corr}, it is evident that the
typical correlation times are much smaller than the average escape times at
all the examined temperatures. Since the correlation time is a reasonable
estimate for the time required to reach a ``local thermal equilibrium", these
results suggest that, whatever the distribution of initial condition used for
the USs is, the system thermalizes before escaping. In fact, USs performed by
starting from different sets of initial conditions lead to close $\Gamma$
values. For instance, in Fig.~\ref{confro-term} one can compare the results of
simulations started from a Maxwellian distribution of the velocities
(dot-dashed line) with those obtained by gradually warming (with a
$7\times10^{-4}$ rate) an initially frozen configuration. The relative
differences are much smaller than the deviations from the theoretical
expectation (dashed line).

\begin{figure}
\cl{
\includegraphics[clip,width=8.truecm]{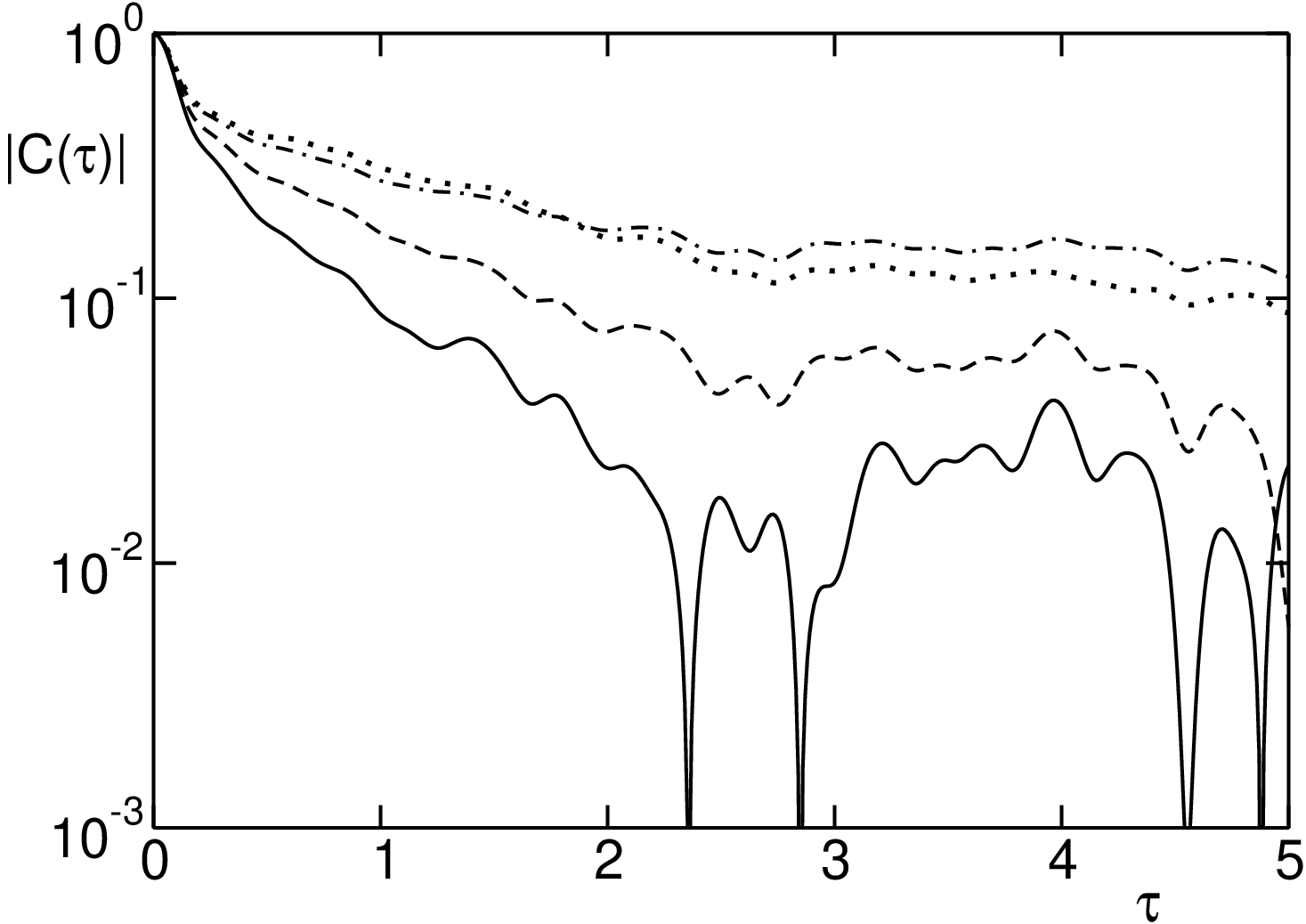}
}
\vskip 0.5 cm
\caption{Initial decay of the absolute value of the autocorrelation function
$C(\tau)$ for the potential energy during canonical simulations
at $T=0.02$ (continuous line), $T=0.04$ (dashed line),
$T=0.06$ (dot dashed line),$T=0.08$ (dotted line).
For each temperature $C(\tau)$ is averaged over a time span of 50 and
over 10 different trajectories.
}
\label{canoT81}
\end{figure}

However, Tab.~\ref{tab-corr} brings forth some interpretative problems: while the
correlation time of the kinetic energy does not depend on $T$ and is
proportional to the friction coefficient, the correlation time of the
potential energy $V(t) = V_1(t) + V_2(t) + V_3(t)$ decreases with temperature.
This phenomenon can be directly observed in Fig.~\ref{canoT81}, where the
absolute value of $C(\tau) = \langle V(t) V(t+\tau) \rangle$ is plotted for
different temperatures. The partial slowing down is an entirely nonlinear
effect, since no temperature dependence can arise in a purely harmonic
potential. In the next subsections we shall see that nonlinearities are
indeed at the origin of the limited validity of the theoretical formula.

\subsection{Role of nonlinearities}

Langer's estimate assumes that the potential is harmonic in the vicinity of
both the NC and the saddle. In order to test whether nonlinear corrections
may be important, we have estimated expression (\ref{erate}) under the
hypothesis of a fully separable potential. In fact, from the products of the
one-dimensional integrals in Eqs.~(\ref{rate_1da},\ref{rate_1db}), one can
at least establish whether nonlinear corrections are truly important. In
practice, we have evaluated the integrals along the eigendirections of the
Hessian in the NC and in the corresponding saddle. The integration interval for
the $i$-th eigendirection is set equal to $[-r^*_i, r^*_i]$, where 
$r_i^* = 3 \sqrt{2\pi T/\omega_a^{(i)}}$ for $I_a$ and analogously $r_i^* = 3
\sqrt{2\pi T/\omega_\perp^{(i)}}$ for $I_\perp$.

\begin{figure}
\cl{
\includegraphics[clip,width=9.truecm]{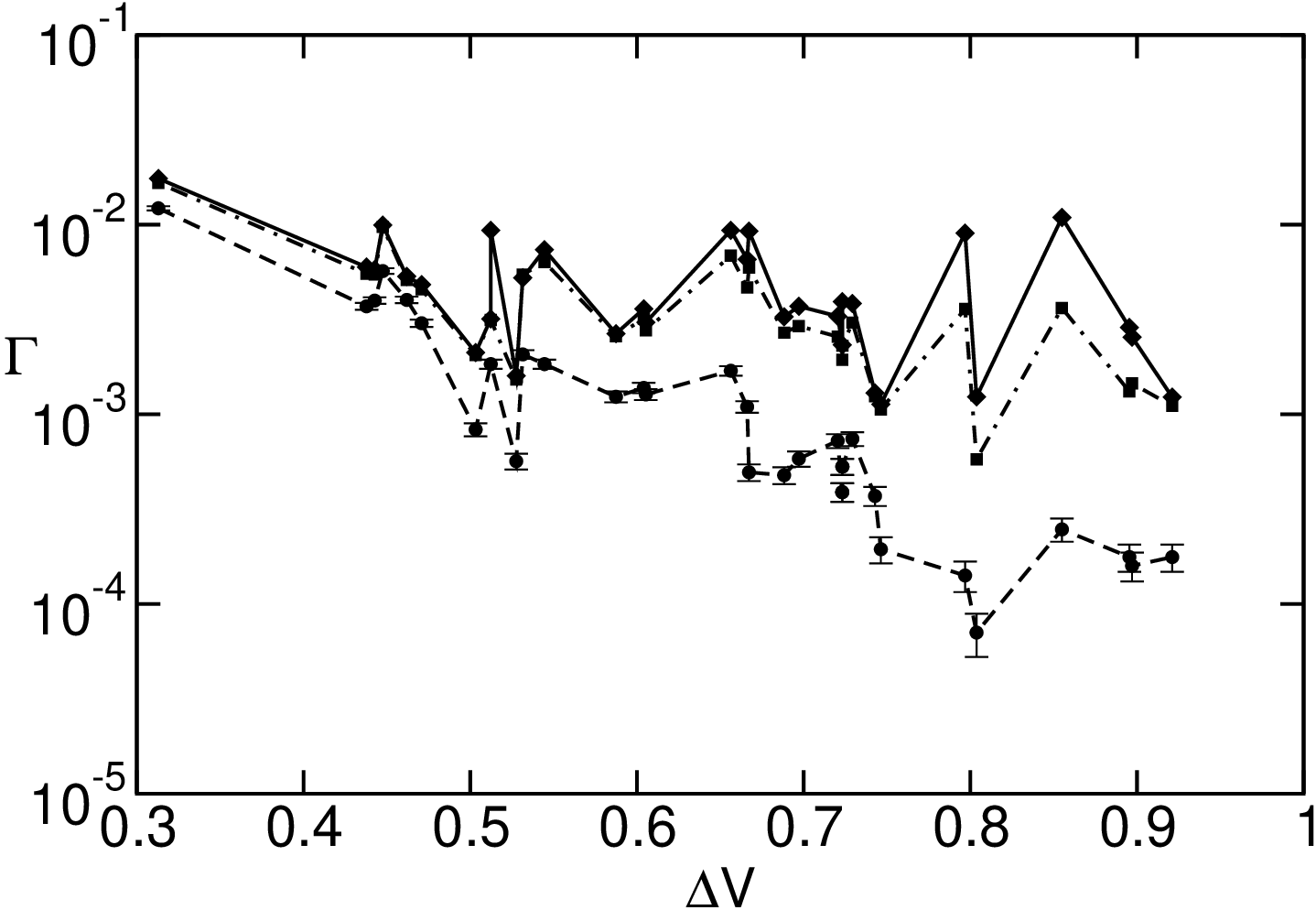}
}
\vskip 0.5 cm
\caption{Transition rate $\Gamma$ versus the barrier height: comparison of
Langer's expression (\ref{rate_n}) (solid line) with numerical estimate 
(\ref{rate_num}) (dashed line) and with expression (\ref{erate}) equipped with
Eqs.~(\ref{rate_1da},\ref{rate_1db}) (dot-dashed line) for S0 at
T=0.1. Numerical estimates have been obtained with a sampling time
$\Delta t = 0.03$ and performing N= 3000 USs.}
\label{1dint}
\end{figure}

The comparison of this expression with both numerical results and the standard
Langer's formula is presented in Fig.~\ref{1dint} for the sequence S0. Although
there is no reason to expect the potential to be separable, it is interesting
to notice that the refined theoretical expression improves over Langer's
formula. On the other hand, the remaining sizeable deviations from the
numerical results indicate the need of a really improved theoretical formula.

\begin{figure}
\cl{
\includegraphics[clip,width=8.truecm]{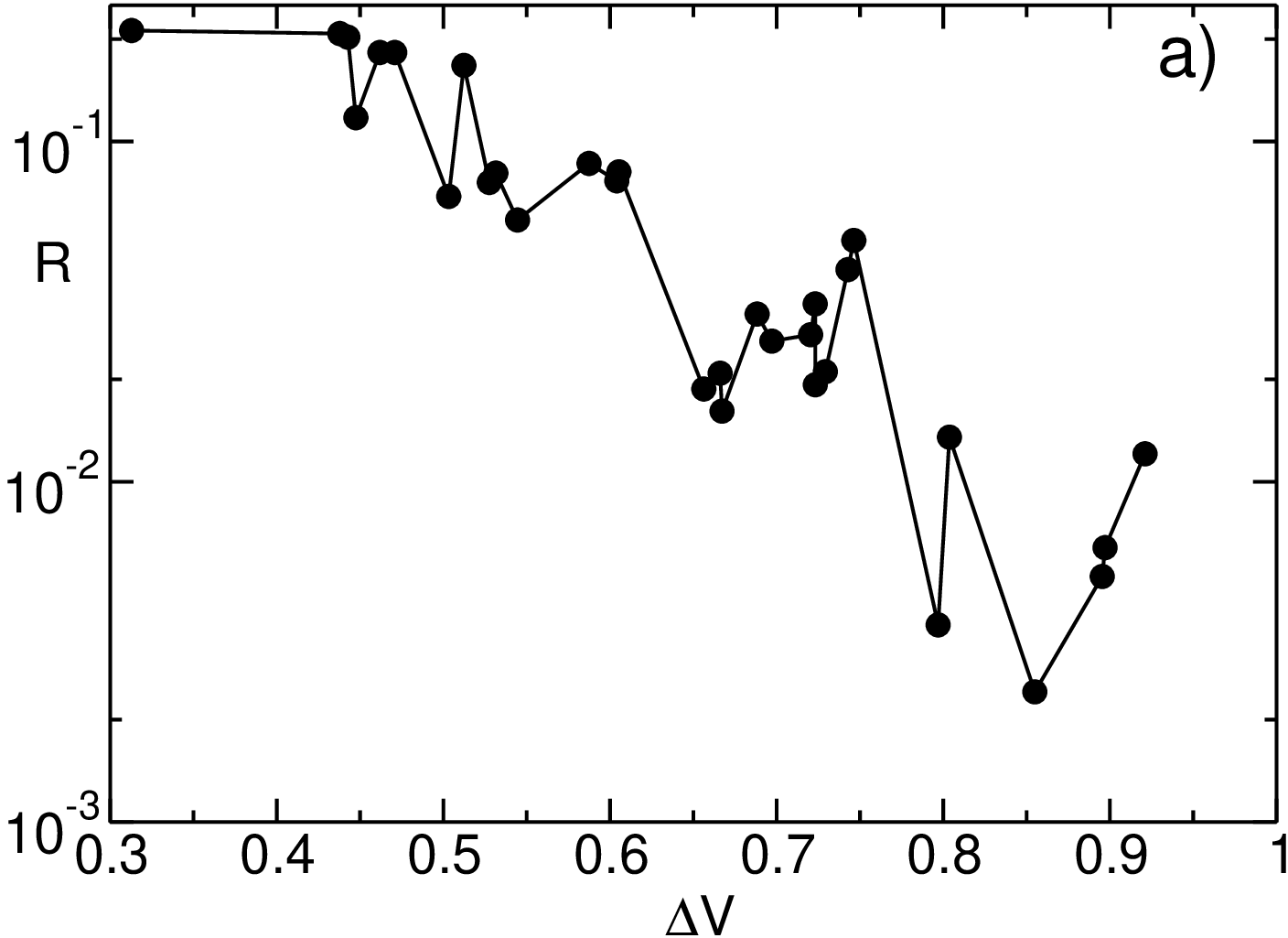}
\includegraphics[clip,width=8.truecm]{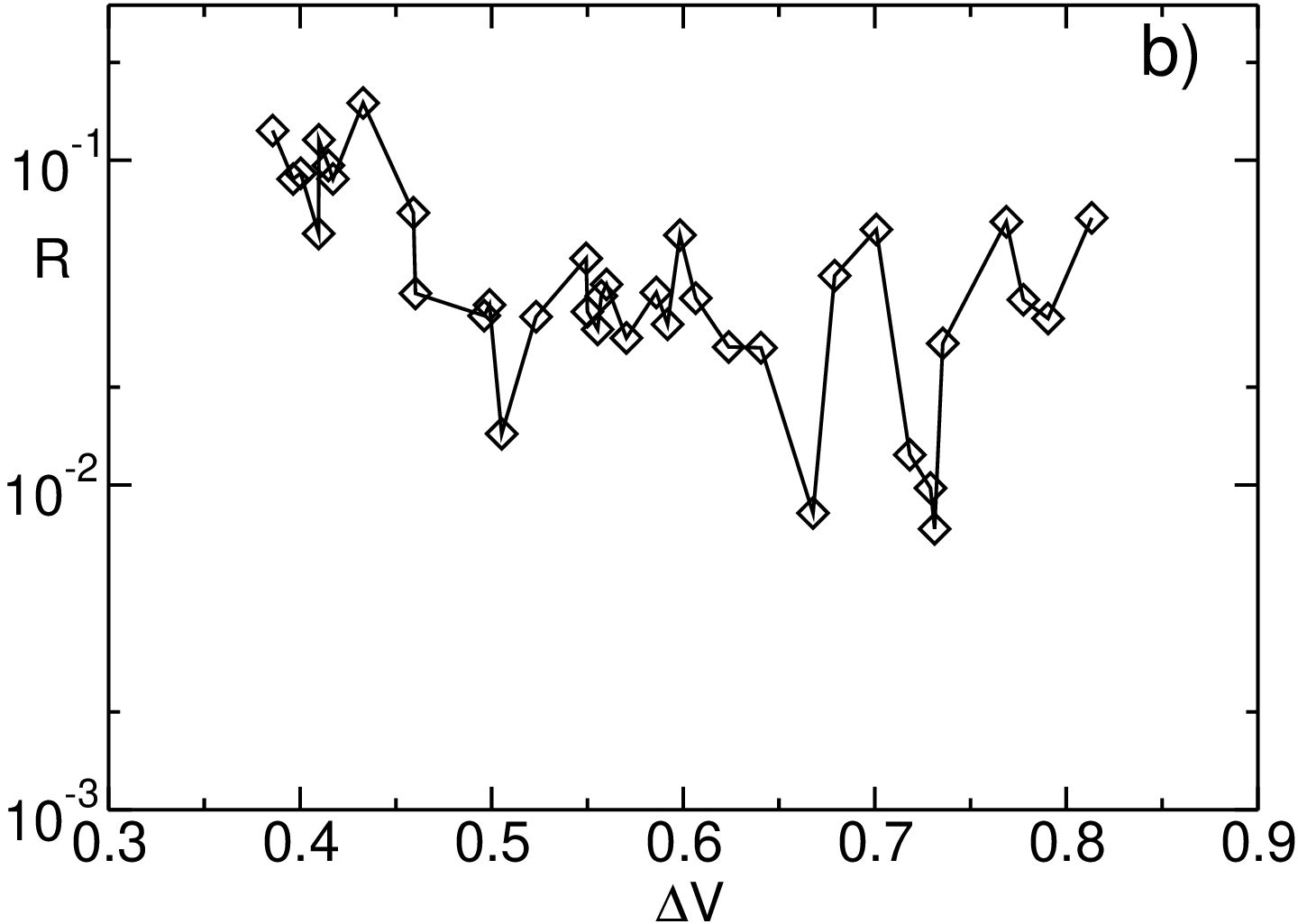}
}
\cl{
\includegraphics[clip,width=8.truecm]{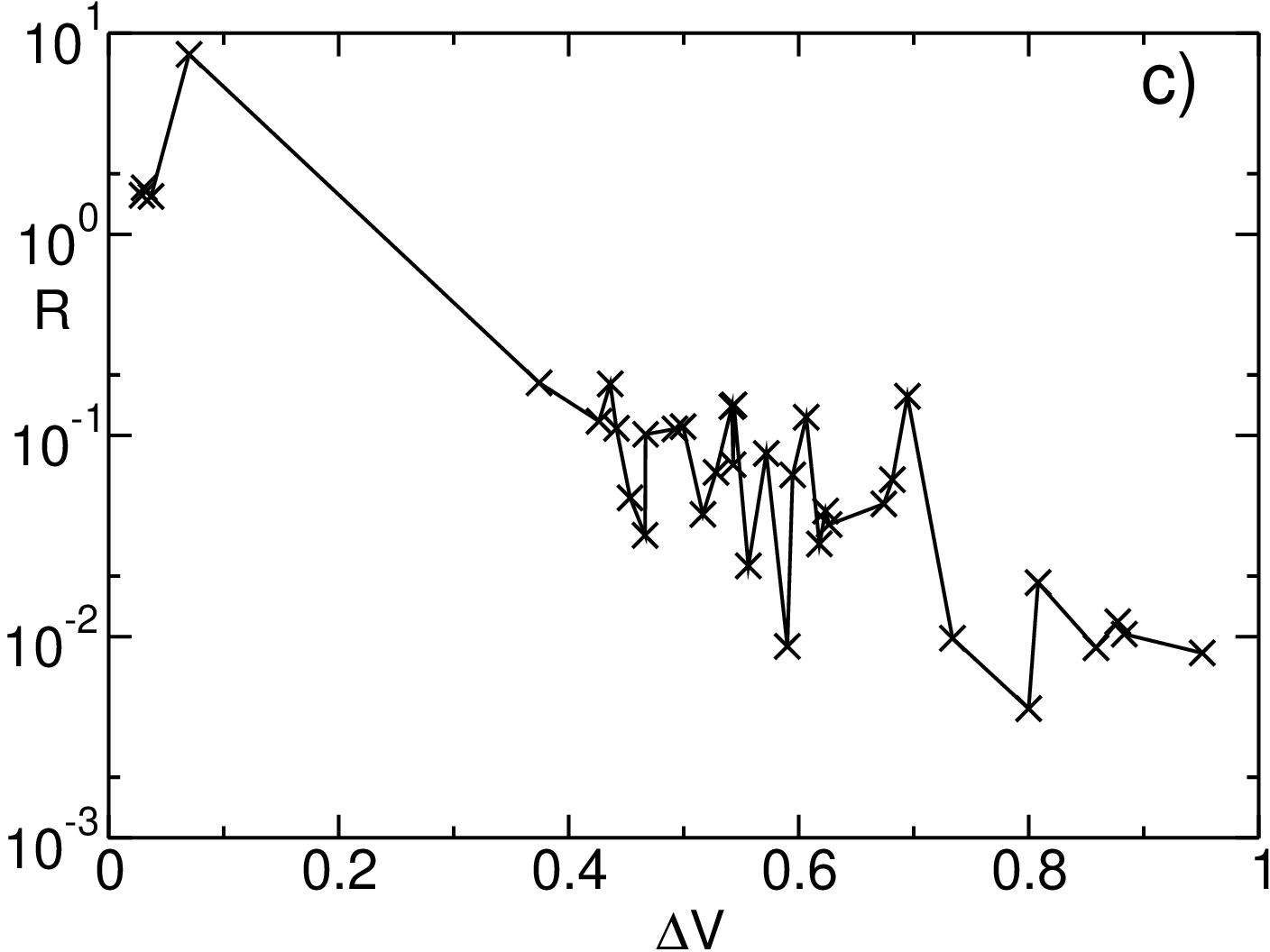}
}
\caption{Entropy ratios $R$ versus the barrier height $\Delta V$,
for the three studied sequences: (a) S0 , (b) S1 and (c) S4.
}
\label{figr}
\end{figure}

The derivation of Langer's formula reveals that an activation process
can be decomposed into an Arrenhius factor, controlled by the energy difference
$\Delta V$, and an entropic factor $R$. The analysis of the latter one conveys
useful information on the structure of the potential energy landscape and helps
sheding some light on the above mentioned discrepancies. While, a priori, there
is no reason to expect $R$ to be either smaller or larger than 1, in practice,
it is almost always smaller than 1 (see Fig.~\ref{figr}); the only exceptions 
are four saddles all around the energy minimum of S4, which are also 
characterized by extremely low barriers. In fact, such saddles are quite
peculiar in that they almost coincide with the NC except for the first 3-4
beads, which are, on the other hand, relatively distant from the core of the
configuration\cite{note3a}. Similar distributions of $R$-values have been also
found in small clusters of particles interacting through Lennard-Jones
potentials \cite{parisi,ruocco}. 

The presence of such large entropic factors accounts for the peculiar dependence
of the escape rate on $\Delta V$ for S4. The abrupt drop of $\Gamma$ when
$\Delta V$ is decreased below 0.2 (see Fig.~\ref{prob})is due the 
smallness of the
entropic contribution ($\Gamma$ is inversely proportional to $R$). This
interpretation is further confirmed by the slow dependence of $\Gamma$ on the
temperature for such saddles.

Leaving aside this peculiarity, there is an average tendency of $R$ to decrease
upon increasing the barrier height in both S0 and S4; this indicates that
higher barriers correspond to flatter saddles and thereby can be more easily
overcome. This is not the case of the good folder, where $R$ does not show any
clear trend and is always bounded in the interval $[10^{-2},10^{-1}]$. 
Accordingly, all the NNMs are entropically equivalent and the escape rate is
essentially determined by the Arrhenius factor. 

\begin{figure}
\cl{
\includegraphics[clip,width=9.truecm]{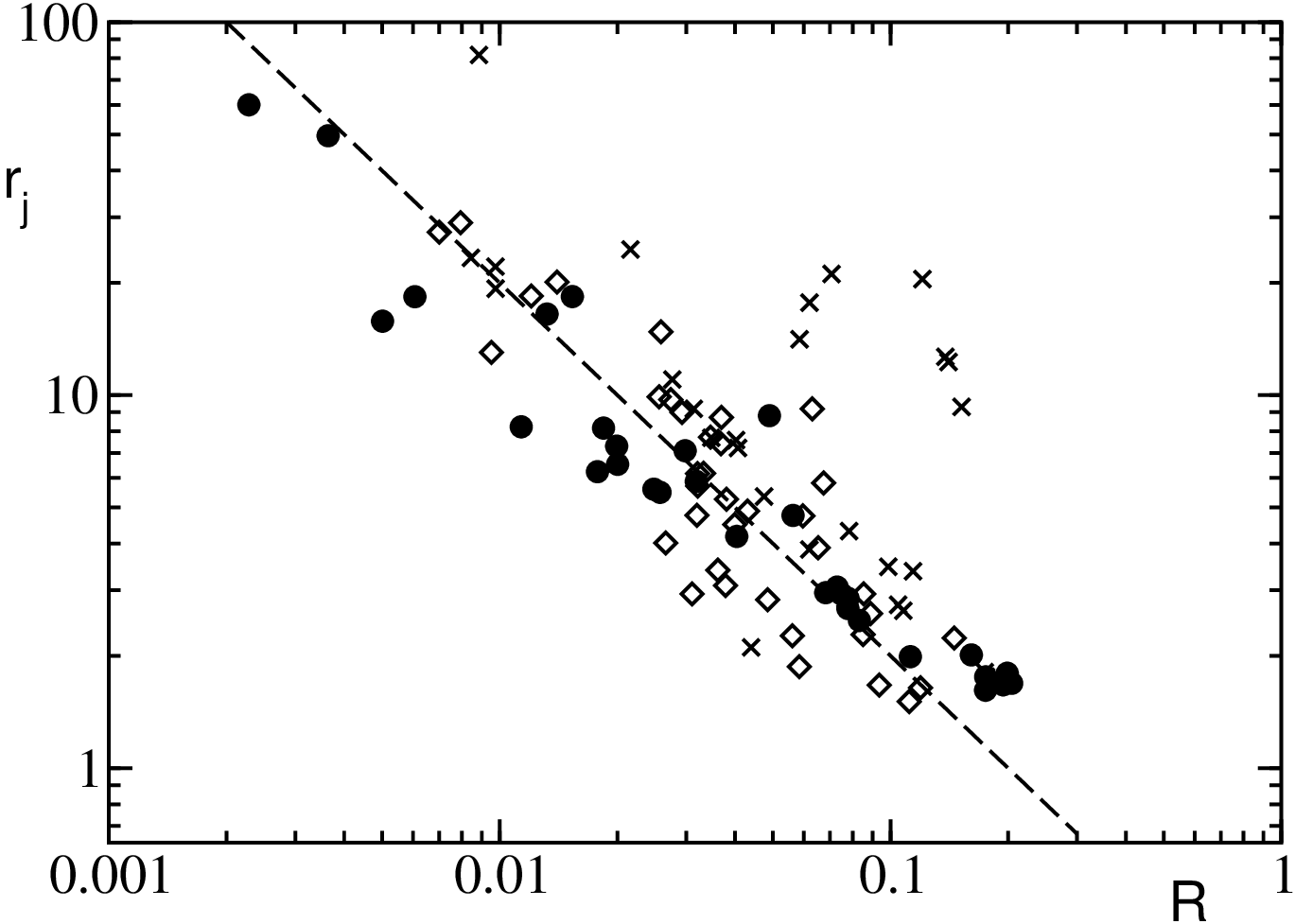}
}
\vskip 0.5 cm
\caption{The relative error $e_j$ versus the entropic factor $R$ for S0 (circles),
S1 (diamonds) and S4 (crosses). 
The data have been obtained by sampling a trajectory
every 0.3 time units at a temperature $T=0.1$.  The dashed
line is a guide for the eye and it corresponds to a
slope one.}
\label{systdev}
\end{figure}

The most interesting observation can be, however, made by parametrizing 
the relative error $e_j$ (see Eq.~(\ref{rel_er})~)~ of the escape rate 
$r_j$ towards the
$j$th NNM. From Fig.~\ref{systdev}, one can see that the deviation of Langer's
formula becomes systematically larger upon decreasing $R$. This is
qualitatively understandable, since a small $R$ implies a flat saddle;
therefore it is reasonable to expect nonlinear corrections to be more relevant.
What is less obvious is the observed dependence of $e_j$ on $R$. The data
reported in Fig.~\ref{systdev} reveal that $e_j \approx a/R$, which amounts to
conjecturing that $\Gamma \simeq 1/(R + a)$ with $a$ basically independent
of $j$. This is quite a remarkable result considering that the fit is more
convicing at larger temperatures, when values as large as 60 of $e_j$ are
observed.

\section{Concluding Remarks}

We have studied in detail the unfolding dynamics of short homo- and
heteropolymeric chains made of $N$-beads in two dimensions. Polymers have
been simulated via an off-lattice model previously introduced~\cite{still} and
their evolution has been examined within the canonical ensemble. Our results
suggest that the dynamics of polymers within their {\it native valley} can be
described as a thermally activated process in a $2 N$-dimensional space in a
whole range of temperatures above and below the folding temperature.

As a matter of fact, Langer's estimate for the escape rate represents a good
approximation for all the examined sequences at low temperatures. We have
verified that  discrepancies between Langer's estimate and numerical data are
mainly due to the poor approximation (limited to the harmonic terms) of
the potential around the stationary points. A better estimate can be derived
by taking into account higher order terms in the expansion of the potential.
Since the folding temperatures for the homopolymer (S0) and the bad folder (S4)
are relatively low, for these sequences  the dynamics within the native valley 
can be reproduced reasonably well already with Langer's approximation. While
for the good folder (S1), this is not the case, thus suggesting a more relevant
role of nonlinearities at the folding temperature. In any case all the examined
estimates turn out to be upper bounds for the escape rates.

An analysis of the entropic contribution to the escape rate suggests that the
folding behaviour of a sequence can be related to topological properties of the
landscape around the NC. For bad folders higher energy barriers are associated
to flatter saddles, thus favouring jumps towards more unfolded configurations.
On the other hand, for the good folder the entropy  ratio seems not to be
related to the heigth  of barriers.

We would like to stress that our analysis amounts to explore the free energy
landscape of a polymer, since in the jumping rate estimates  are included not
only Kramers' terms but also entropic contributions. The relevance of the
latters in determining equilibrium  and kinetic properties of peptides  has
been recently pointed out in  \cite{krivov,evans}).

In order to further explore the role of activation processes for the complete
folding dynamics  we plan to extend our  analysis to the whole energy
landscape. A complete graph  describing all the folding/unfolding paths with
their associated probabilities will allow to determine equilibrium  properties
of the system and possibly to distinguish bad and good folders.

\acknowledgments
We warmly acknowledge the active contribution given at the early stage of this
work by Annalisa Tiberio\cite{tiberio}. We are also grateful to CINECA in
Bologna and INFM  for providing us access to the parallel computer CRAY T3E 
and to the Beowulf Linux-cluster under the   grant ``Iniziativa Calcolo
Parallelo''. This work has been partially supported under the FIRB project
RBAU01BZJX  ``Dynamical and statistical analysis of biological microsystems''.

\begin{appendix}

\section{Algorithm for the identification of the saddles}

The algorithm described in this appendix (see also ~\cite{tiberio}~) aims at
determining the path of minimal potential energy connecting two minima indicated
as ${\bf a}_1$ and ${\bf a}_2$.

Due to the symmetries of the potentials defined in Section II
(see eq. (\ref{hamil})~), each spatial configuration of the polymeric
chain is defined apart from a translation, a reflection and a rotation around
an axis perpendicular to the $xy$-plane. Accordingly, after having expressed
${\bf a}_1$ in an arbitrary reference frame, it is convenient to determine the
coordinates of ${\bf a}_2$ are by minimizing its Euclidean distance from
${\bf a}_1$ with respect to the above mentioned symmetry transformations. 

The algorithm then consists in evolving a suitably chosen path connecting
${\bf a}_1$ and ${\bf a}_2$ according to a gradient dynamics until the maximum
of the energy along the path converges to a minimum corresponding to a
saddle. More precisely, the approach is split into three steps:

\begin{enumerate}

\item{ \bf Choice of the initial configurations}
The initial path ${\mathcal{C}}_0$ connecting $\bf{a_1}$ and $\bf{a_2}$ is
generally chosen by linearly interpolating between their coordinates,
\begin{eqnarray}
x(i)=x^1(i)+r(x^2(i)-x^1(i)) \nonumber \\
y(i)=y^1(i)+r(y^2(i)-y^1(i)) \nonumber  \\
i=1, \ldots , N \quad .
\label{ck0}
\end{eqnarray}
The sequence of initial configurations along the path ${\mathcal{C}}_0$
is fixed by varying the parameter $\emph{r}$ between 0 and 1 (we typically
choose $\emph{r} = m/100$, $0 \le m \le 100$). 

\item{\bf Evolution of the configurations}
Each configuration is then let evolve according to the gradient dynamics
\begin{eqnarray}
{\dot x}_i = - \frac{1}{\tilde{\gamma}}
\frac{\partial H}{\partial x_i} \quad \quad \nonumber \\
{\dot y}_i = - \frac{1}{\tilde{\gamma}} \frac{\partial H}{\partial y_i}
\quad \quad .
\label{over}
\end{eqnarray}
In practice, the damping coefficient ${\tilde \gamma}$ can be chosen equal to
one, since it only determines the evolution time scale. The integration
time step $\delta t$ is adapted to the instantaneous value of the force field,
$\delta t = \min (.01,.01/F_{max})$, where 
$F_{max} = \max_i \{ |f^x_i|, |f^y_i|\}$ while $f^x_i$ and $f^y_i$ are the $x$
and $y$ components of the force acting on the $i$th bead.

\item{ \bf Interpolation phase}
After letting evolve the system for a time $\bar t = 10 \delta t$, the Euclidean
distance $\Delta_{m}(\bar t)$ between the $m$th and the $(m+1)$st point is
computed. If $\Delta_m(\bar t) > 2 \Delta_m(0)$, a new configuration,
is added between the two points by linearly interpolating between them.
If $\Delta_{m}(\bar t) < \Delta_m(0)/2$ the $(m+1)$-st configuration is removed.
In this way, we are able to work with a set of uniformly distributed
configurations, without loosing resolution in the regions where the energy
gradient is large.

The last two steps are repeated until $F_{max}$ in the maximum energy
configuration along the path becomes smaller than a fixed threshold typically
chosen equal to $10^{-3}$. The coordinates of the saddle point are finally
refined by implementing a standard Newton scheme.
\end{enumerate}

\begin{figure}[ht]
\cl{
\includegraphics[clip,width=7.truecm]{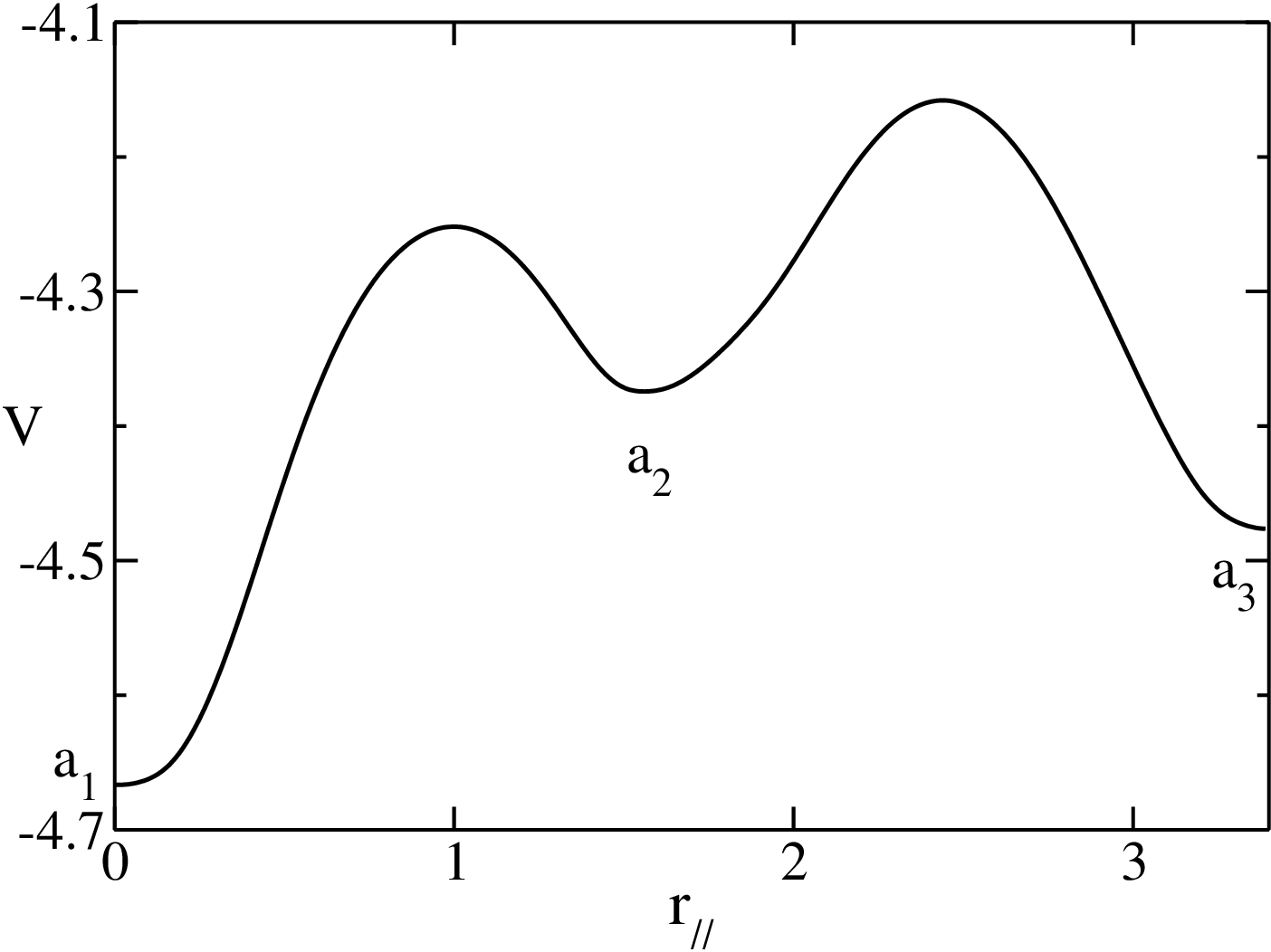}
}
\vskip 0.5 cm
\caption{Potential energy profile $V$ connecting the NC
${\bf a_1}$ to a 2NNM ${\bf a_3}$
via 2 saddles for the sequence S1, the symbol ${\bf a_2}$
indicates a NNM. The index
$r_\parallel$ refers to the distance measured
along the unstable manifold connecting the minima.
}
\label{gobbe}
\end{figure}

On one hand, the path connecting two generic minima can exhibit more than one
relative energy maximum (see, e.g., Fig.~\ref{gobbe}): this is an indication that
the two minima are not nearest neighbours. In this case, our approach allows
identifying new minima. On the other hand, it can happen that neighbouring
minima are separated by more than one saddle: upon choosing different initial
paths, one can, in principle, identify all saddles. Our simulations suggest
that multiple saddles are not very common at least in the vicinity of the NC.

\end{appendix}



\begin{thebibliography}{100}
\bibitem{note1}
For the sake of simplicity, all variables are expressed in
adimensional units. The harmonic term turns out to be stiff because $\alpha=20$
is much larger than the constants corresponding to the two other interaction
processes included in the model.
\bibitem{note2}
The original formula differs by a factor-2 since we are here referring
to the first passage time at the boundary, rather than to the neighboring
minimum.
\bibitem{note3}
An upper bound for the lifetime $\tau_l$ of a NNM is given by the inverse of the
escape rate from it to the NC evaluated by employing Eq.~(\ref{rate_n}). 
With reference to temperatures in the interval $[0.02,0.08]$, we have found
$12 \le \tau_l \le 37$ for S0, $7 \le \tau_l \le 20$ for S1, and 
$0.06 \le \tau_l \le 1.06$ for S4.
\bibitem{note3a} Notice that the first three beads of S4 are all polar and are
thus repelled by the vast majority of the remaining hydrophobic ones.
\end{thebibliography}
\end{document}